\documentclass[
 twocolumn,
 amsmath,
 amssymb,
 aps,
 prb,
]{revtex4-2} 
\usepackage{xcolor}
\usepackage{graphicx}
\usepackage{dcolumn}
\usepackage{multirow}
\usepackage{adjustbox}
\usepackage{bm}
\usepackage[utf8x]{inputenc}
\usepackage{pslatex}
\usepackage{textcomp} 
\usepackage{hyperref}
\usepackage[T1]{fontenc} 
\usepackage{lmodern} 
\usepackage{breakurl} 
\usepackage{subfig} 
\usepackage{booktabs}
\graphicspath{{./figures/}}
\usepackage{dcolumn}
\usepackage{tabularx}
\usepackage{dcolumn} 
  \newcolumntype{d}[1]{D{.}{.}{#1}} 
\usepackage{siunitx} 
\begin{document}
\begin{sloppypar} 

\title{Effect of transition metal doping on magnetic hardness of CeFe$_{12}$-based compounds}

\author{Justyn Snarski-Adamski}
\email [Corresponding author: Justyn Snarski-Adamski\\] {justyn.snarski-adamski@ifmpan.poznan.pl}

\author{Mirosław Werwiński}

\address{Institute of Molecular Physics, Polish Academy of Sciences, M. Smoluchowskiego 17, 60-179 Poznań, Poland}

\begin{abstract}
ThMn$_{12}$-type ternary cerium alloys with tetragonal structure (Pearson symbol tI26, space group $I$4/$mmm$) are considered as promising  materials for permanent magnets.
In this work, compositions of CeFe$_{11}$X (s.g. $Pmmn$, No.~59) and CeFe$_{10}$X$_2$ (s.g. $P$4/$mmm$, No.~123) with all 3$d$, 4$d$, and 5$d$ transition metal substitutions are considered. 
Since many previous studies have focused on the CeFe$_{11}$Ti compound, this particular compound became the starting point of our considerations and we gave it special attention.
We first determined the optimal symmetry of the simplest CeFe$_{11}$Ti structure model.
We then observed that the calculated magnetocrystalline anisotropy energy (MAE) correlates with the magnetic moment, which in turn strongly depends on the choice of the exchange-correlation potential.
MAE, magnetic moments, and magnetic hardness were determined for all compositions considered.
Moreover, the calculated dependence of the MAE on the spin magnetic moment allowed us to predict the upper limits of the MAE. We also showed that it does not depend on the choice of the exchange-correlation potential form.
The economically justifiable compositions with the highest magnetic hardness values are CeFe$_{11}$W, CeFe$_{10}$W$_{2}$, CeFe$_{11}$Mn, CeFe$_{10}$Mn$_{2}$, CeFe$_{11}$Mo, CeFe$_{10}$Mo$_{2}$, and CeFe$_{10}$Nb$_{2}$.
However, calculations suggest that, like CeFe$_{12}$, these compounds are not chemically stable and could require additional treatments to stabilize the composition.
Further alloying of the selected compositions with elements embedded in interstitial positions confirms the positive effect of such dopants on hard magnetic properties.
Subsequent calculations performed for comparison for selected isostructural La-based compounds lead to similar MAE results as for Ce-based compounds, suggesting a secondary effect of 4$f$ electrons. Our preliminary results obtained using the intra-atomic Hubbard repulsion term  showed a relatively small difference for CeFe$_{12}$ compared to the results without this correction.
Calculations were performed using the full-potential local-orbital electronic structure code FPLO18, whose unique fully relativistic implementation of the fixed spin moment method allowed us to calculate the MAE dependence of the magnetic moment.

\end{abstract}

\maketitle

\section{Introduction}

%
Rare-earth permanent magnets are used in industries such as automotive, aerospace, electronics, and renewable energy.
They are alloys consisting mainly of transition metals and rare-earth elements.
They have remarkable magnetic properties, such as very high energy density |BH|$_{max}$ and the ability to operate at high temperatures. 
Of particular interest are rare-earth compounds such as Nd$_{2}$Fe$_{14}$B \cite{toga_anisotropy_2018}
and SmCo$_{5}$ \cite{das_anisotropy_2019},
which exhibit extremely high values of magnetocrystalline anisotropy.
However, the high volatility of rare-earth prices, which became clearly evident during the so-called  {\it rare-earth crisis} in 2011 \citep{bourzac_rare-earth_2011}, has mobilized the international research community to search for a new generation of hard magnetic materials with reduced rare-earth content~\cite{niarchos_toward_2015, hirosawa_current_2015, li_recent_2015, hirosawa_perspectives_2017, skomski_magnetic_2016, ener_twins_2021, werwinski_ab_2017-1}. 
This search, although mainly performed by experimental methods, is complemented by groups conducting first-principles calculations of intrinsic properties and magnetic simulations at the microstructural level.

%
In the ongoing search for new permanent magnets, special attention is being given to magnetic compounds with ThMn$_{12}$-type structures based on Nd, Sm, but also less expensive Ce~\cite{gutfleisch_magnetic_2011, delange_crystal-field_2017,sozen_ab_2019, zhou_magnetic_2014, delange_crystal-field_2017, hadjipanayis_thmn12-type_2020}. 
For example, SmFe$_{11}$Ti and SmFe$_{11}$V systems have been found to exhibit uniaxial magnetocrystalline anisotropy~\cite{hadjipanayis_thmn12-type_2020}. 
Whereas, the Curie temperature ($T_{C}$) measured for SmFe$_{11}$V is equal to 634~K, the saturation  magnetization is 11.2~kG (1.12~T), and the anisotropy field is 8.7 MA\,m$^{-1}$ \citep{hadjipanayis_thmn12-type_2020}.
Among Ce compounds with ThMn$_{12}$-type structure, CeFe$_{11}$Ti and CeFe$_{12-x}$Mo$_{x}$ are particularly interesting. 
The experimental magnetocrystalline anisotropy energy (MAE) measured at 300~K for CeFe$_{11}$Ti range from 0.62 to 1.1~MJ\,m$^{-3}$ \cite{pan_structural_1994, akayama_physical_1994}, while the MAE measured at 1.5~K is 1.78~MJ\,m$^{-3}$ ($H_a$~=~35~kOe) \citep{pan_structural_1994}. The magnetic moment ($m$) measured experimentally at 1.5~K is equal to 18.62~$\mu_B$\,f.u.$^{-1}$ ($\sigma_S$~=~129.6 emu\,g$^{-1}$) \cite{pan_structural_1994}, and at 4.5~K is equal to 17.46~$\mu_B$\,f.u.$^{-1}$ \citep{isnard_hydrogen_1998}, 23.14~$\mu_B$\,f.u.$^{-1}$ ($\sigma_S$~=~1.55~T) \citep{akayama_physical_1994}, while $T_{C}$ is equal to 487~K \citep{isnard_hydrogen_1998, pan_structural_1994}.
Example values of magnetocrystalline anisotropy energy for CeFe$_{11}$Ti obtained from first-principles calculations are equal to 1.19, 1.50, 1.57, and 1.98~MJ\,m$^{-3}$~\cite{akayama_physical_1994,ke_intrinsic_2016-1, martinez-casado_cefe11ti_2019}. 
For CeFe$_{11}$Ti, the formation of a hard magnetic phase between the range from 700 and 1100~K has also been both theoretically and experimentally determined \cite{sozen_ab_2019, maccari_correlating_2021}.
Furthermore, experimental analysis of Mo concentration in CeFe$_{12-y}$Mo$_{y}$, where $y$~=~1, 1.5, and 2, showed that the $T_{C}$ for CeFe$_{11}$Mo is 421~K and decreases with Mo content, while for the nitrided counterpart the $T_{C}$ is 643~K~\cite{zhou_magnetic_2014}. 
|BH|$_{max}$ takes the highest value for CeFe$_{11}$Mo equal to 2.39~kJ\,m$^{-3}$ after annealing \citep{zhou_magnetic_2014}. 

\begin{figure*}[t]
\centering
\subfloat[CeFe$_{11}$Ti -- $Pmmn$]{\label{fig:Pmmn}
\includegraphics[clip, width=0.33\textwidth]{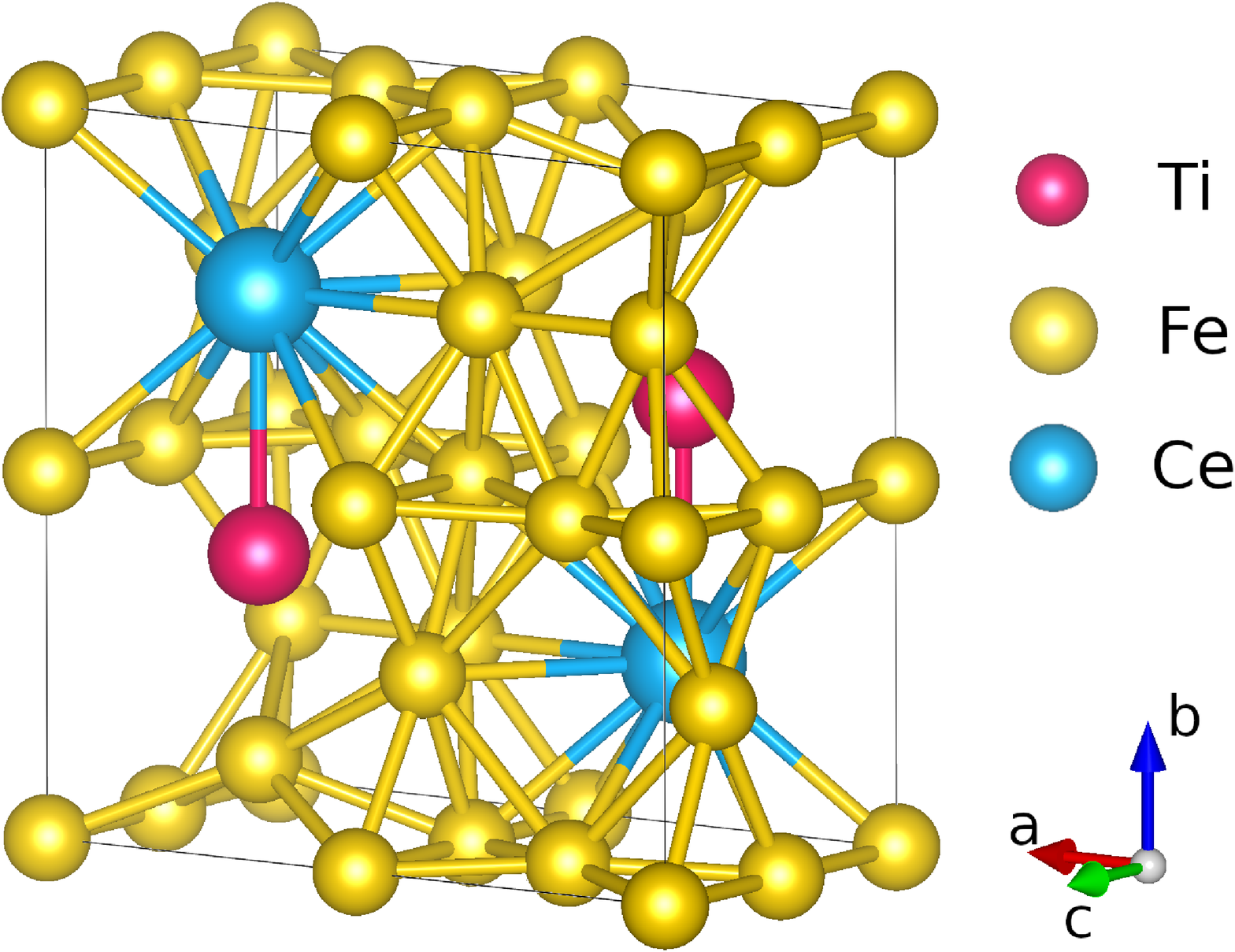}}
\hfill
\subfloat[CeFe$_{10}$Ti$_{2}$ -- $P$4/$mmm$]{\label{fig:CeFe10W2_P4mmm}
\includegraphics[clip, width=0.23\textwidth]{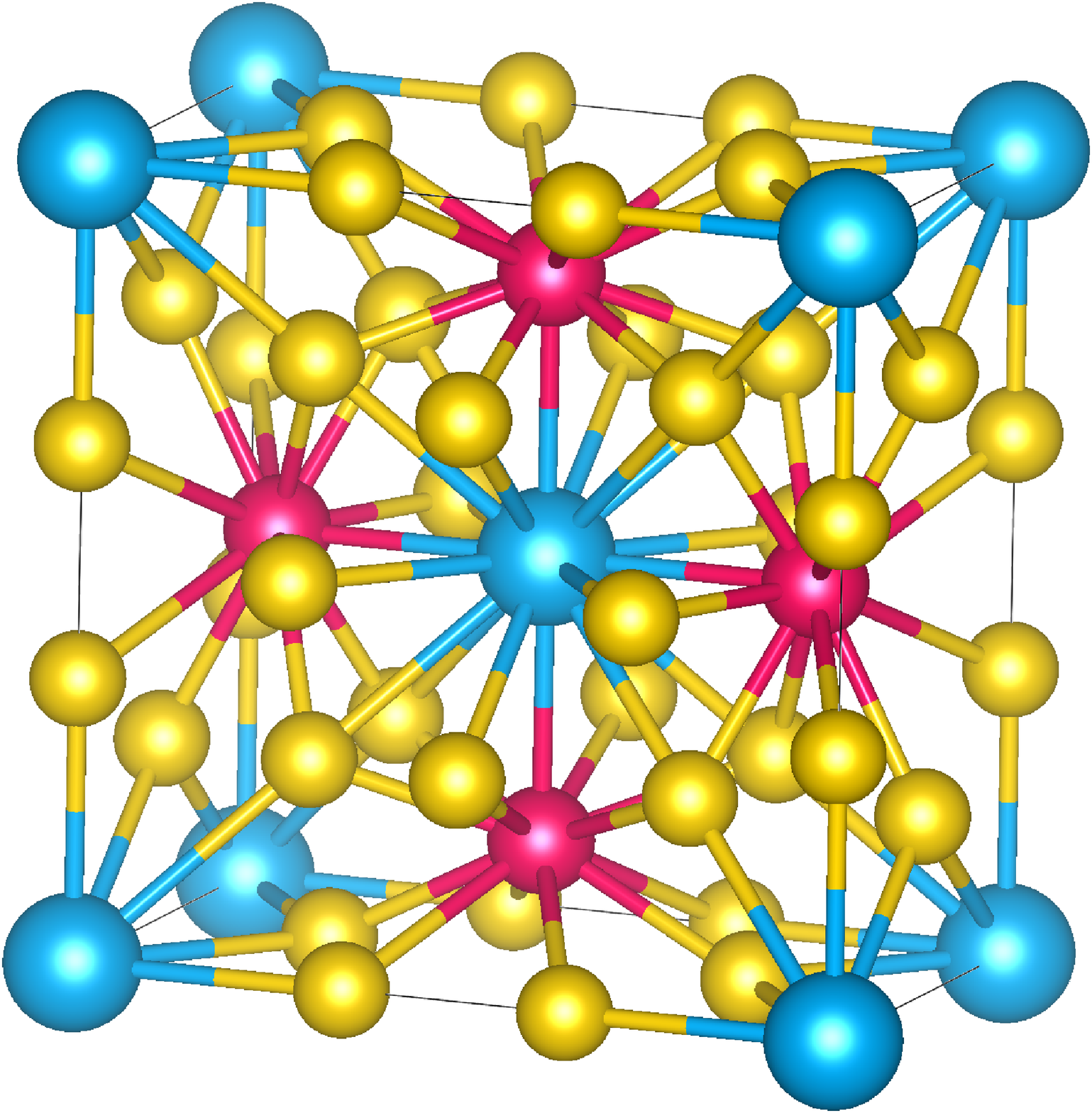}}
\hspace{8mm}
\subfloat[CeFe$_{12}$ -- $I$4/$mmm$]{\label{fig_struct}
\includegraphics[clip, width=0.36\textwidth]{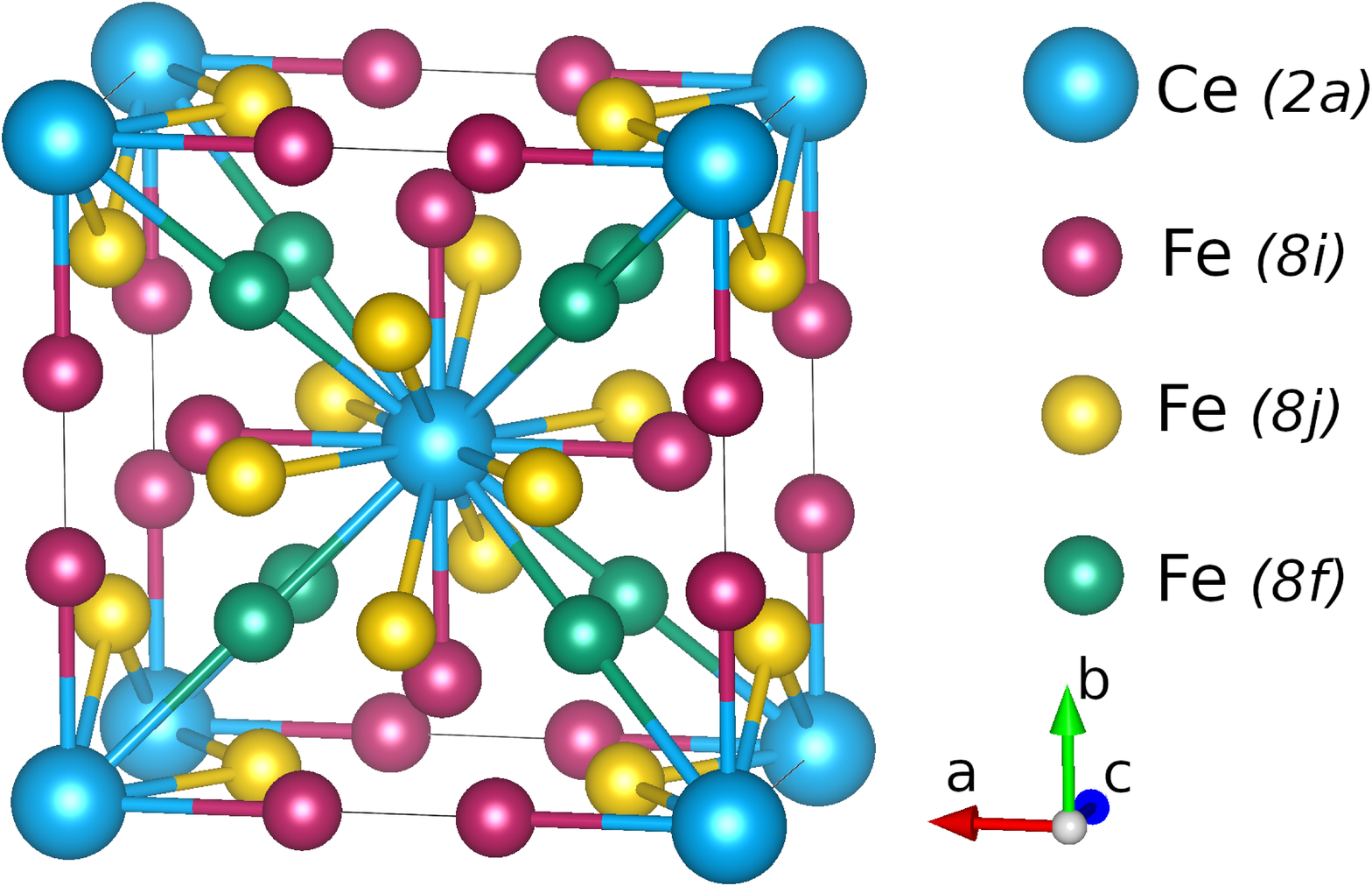}}
\caption{
Crystal structure models of 
(a) 
CeFe$_{11}$Ti (s.g. $Pmmn$, No.~59), 
(b) 
CeFe$_{10}$Ti$_{2}$ (s.g. $P$4/$mmm$, No.~123), and
(c) 
CeFe$_{12}$ (s.g. $I$4/$mmm$, No.~139).
CeFe$_{12}$ structure is a superstructure of the ThMn$_{12}$ type with 2$a$, 8$i$, 8$j$, and 8$f$ Wyckoff positions.
The lattice parameters of all models ($a$~=~8.539, $c$~=~4.780~\AA{}) were adopted from CeFe$_{11}$Ti~\cite{isnard_hydrogen_1998}. 
}
\end{figure*}

%
The aim of this study is to theoretically determine the amount and type of dopants in CeFe$_{12}$-based alloys that will lead to maximizing their magnetic hardness.
The CeFe$_{11}$X and CeFe$_{10}$X$_2$ compositions (X~=~3$d$, 4$d$, 5$d$ elements) and, for comparison, the thermodynamically unstable CeFe$_{12}$ compound will be considered.

%
For rare-earth materials, the simplest theoretical  approaches (local-density approximations (LDA) / generalized gradient approximation (GGA)) based on density functional theory (DFT) are often insufficient. Even simple rare earth metals, such as $\alpha$ and $\gamma$ Ce phases, have recently been extensively studied to identify optimal methods that lead to credible results \citep{tran_nonmagnetic_2014}. Although it is known that LDA/GGA alone usually does not reproduce the properties of Ce compounds well, in this work we chose to use GGA specifically. There are several factors behind our choice. (1) Our initial calculations going beyond GGA using the intra-atomic Hubbard repulsion term U (GGA + U) showed for CeFe$_{12}$ relatively good agreement with the GGA results. (2) Complementary calculations for isostructural systems with La (not containing $f$ electrons) in place of Ce, confirmed the dominant effect of Fe on the magnetic properties of the considered compounds. (3) Remaining with GGA, we avoided the computational difficulties associated with more advanced approaches, such as multiple magnetic solutions, and effectively analyzed the MAE dependence on the spin magnetic moment together with the effect of doping on the magnetic properties of CeFe$_{12}$-based compounds. Recently we also presented a detailed analysis of the effect of the Hubbard U correction on the electronic configuration of Ce in work on the Ce$_{1-x}$Pr$_{x}$CoGe$_{3}$ system \citep{skokowski_influence_2020}.

Theoretically predicted intrinsic magnetic properties such as magnetization and magnetocrystalline anisotropy energy (MAE) can be used to estimate the upper limits of coercivity and energy product (|BH|$_{max}$) \citep{coey_permanent_2012, korner_theoretical_2016}. The upper coercivity limit is determined by the anisotropy field  defined by the formula  $H_{a}$~=~2 $K_{1}$/($\mu_0$ $M$), where $\mu_0$ is the free space permeability, $M$ is the magnetization, and $K_{1}$ is the anisotropy constant - often approximated as the calculated MAE. In turn, the upper limit of the energy product can be estimated from the formula |BH|$_{max}$ = $M^2$/(4 $\mu_{0}$).

%
While hard magnetic materials typically operate at room temperature and above, the DFT results presented here are calculated for a temperature of 0~K. Therefore, the magnetic moments, magnetocrystalline anisotropy energies, and magnetic hardnesses determined from the calculations should be seen as their upper limits, which typically decrease with increasing temperature.

%
\section{Computational details}

\begin{table}
\caption{\label{tab:CeFe12-struct.param.} Structural parameters of CeFe$_{12}$ (s.g. $I$4/$mmm$, No. 139). In our calculations for the CeFe$_{12}$ compound, we used the lattice parameters measured for CeFe$_{11}$Ti, which are $a$~=~8.539 and $c$~=~4.78~\AA{}, see Ref.~\cite{isnard_hydrogen_1998}.}
\begin{tabular}{p{2.0cm}p{1.5cm}p{1.5cm}p{1.5cm}}
 \hline
 \hline	
Site        &   $x$   	&   $y$	&   $z$\\
\hline				
Ce (2$a$)	&	0		&	0	&	0	\\
Fe (8$i$)	&	0.360	&	0	&	0	\\
Fe (8$j$)	&	0.265	&	0.5	&	0	\\
Fe (8$f$)	&	0.25	&	0.25&	0.25	\\
\hline
\hline
\end{tabular}
\end{table}

Calculations were performed using the full-potential local-orbital electronic structure code FPLO18 with a fixed atomic-like basis set \cite{koepernik_full-potential_1999-1}. The supercell method was used to model all CeFe$_{11}$X and CeFe$_{10}$X$_{2}$ systems (X~=~3$d$, 4$d$, and 5$d$ elements). The generalized gradient approximation (GGA) was used in the Perdew-Burke-Ernzerhof (PBE) form~\cite{perdew_generalized_1996-1}. The local spin density approximation (LSDA) in the forms of von Barth and Hedin (BH)~\cite{barth_local_1972}, Perdew and Zunger (PZ)~\cite{perdew_self-interaction_1981}, and Perdew and Wang (PW92)~\cite{perdew_accurate_1992-1} were also used to analyze the effect of the exchange-correlation potential form on the magnetocrystalline anisotropy energy (MAE).
A 12$\times$12$\times$12 k-mesh was found to lead to well converged MAE results. The density convergence criterion was set to 10$^{-6}$. In all cases, we optimized the geometry with a force convergence criterion of 10$^{-3}$~eV\,\AA{}$^{-1}$.
A similar systematic supercell approach has been previously applied to model (Fe,Co,X)$_2$B and (Fe,Co,X)$_5$PB$_2$ alloys~\cite{edstrom_magnetic_2015-2, werwinski_magnetocrystalline_2018}.

In addition to the more general rule indicating the choice of PBE among the PBE/LDA exchange-correlation potentials for Fe bcc and Fe compounds with transition metals  \citep{kormann_pressure_2009, matyunina_ab_2018, romero_one_2018}, the comparison of magnetic moments obtained by different approximations with experimentally obtained values should be another criterion for making an optimal choice. In the present work the choice of PBE appears to be optimal. However, it should be kept in mind that the results presented in this work, apart from the limitations related to performing calculations for the ground state (0~K), using the supercell method to model chemical disorder, and using a single experimental lattice parameter to model different compounds, may also be affected by the error resulting from the nature of the PBE approximation, which tends to slightly overestimate the obtained values of magnetic moments, which, as we will show, may have some influence on the MAE values obtained.

CeFe$_{12}$ is expected to crystallize in a tetragonal structure with space group {\it I}4/{\it mmm}~\cite{zhou_magnetic_2014, hadjipanayis_thmn12-type_2020, lee_large_2021}, see Table~\ref{tab:CeFe12-struct.param.} and Fig.~\ref{fig_struct}.
For all considered models, we assumed lattice parameters as measured for CeFe$_{11}$Ti~\cite{isnard_hydrogen_1998}, while the Wyckoff positions were optimized using a spin-polarized scalar-relativistic approach. 
For CeFe$_{12}$, we optimize the atomic positions using various exchange-correlation functionals available in FPLO. For CeFe$_{11}$Ti, we considered some of the simplest supercell models with an ideal arrangement of dopant atoms and determined the structure with the lowest total energy ($Pmmn$), see Fig.~\ref{fig:Pmmn}. In the case of  CeFe$_{10}$Ti$_{2}$ with Ti atoms at 8$i$ sites, there is only one possible crystallographic configuration, with a space group $P$4/$mmm$ (No. 123), see Fig.~\ref{fig:CeFe10W2_P4mmm}. In the case of the presented calculations for "1-11-1" systems doped with light atoms such as C, H, B, and N, we assumed that they occupy 2$b$ sites.

For the "1-10-2" tetragonal system, the MAE was determined as the difference between the fully relativistic total energies calculated for the quantization axes [100] and [001].
For the "1-11-1" system, the MAE was evaluated as the difference between the energies calculated for the orthogonal axes [101] and [010]. We chose the unconventional [101] axis because the space group under consideration ($Pmmn$, No. 59) is orthorombic. We interpret the energy value obtained for [101] axis as the value averaged between axes [100] and [001].  In the adopted sign convention, the positive sign of the MAE is consistent with an easy axis of magnetization along the [010] direction, and the negative sign is consistent with in-plane anisotropy. Furthermore, we determined the MAE values from a single iteration of the fully-relativistic approach.
Using the fully relativistic fixed spin moment scheme~\cite{schwarz_itinerant_1984}, we  also study the MAE as a function of the total spin magnetic moment ($m_S$).
Tables with the atomic positions of the investigated compounds can be found in the Appendix section.
The VESTA code was used to visualize the crystal structures~\cite{momma_vesta_2008}.

%
Another important parameter in the context of permanent magnets is the magnetic hardness, defined as:
\begin{equation}
\kappa = \sqrt{\frac{|K|}{\mu_0 M_S^2}}, 
\end{equation}
\noindent
where $K$ is the magnetocrystalline anisotropy constant, $M_S$ is the saturation magnetization, and $\mu_{0}$ is the vacuum permeability. The empirical rule $\kappa$~>~1 specifies  whether the material will resist self-demagnetization~\cite{skomski_magnetic_2016}.
In determining the theoretical value of  $\kappa$, we assume that the anisotropy constant $K$ is equal to MAE, and M$_S$ is evaluated from the calculated total magnetic moment and unit cell volume.
It is worth noting that the magnetic hardness makes sense only for materials that are proper permanent magnets, and thus have a easy-axis of magnetization rather than an easy-plane. Thus, when MAE~<~0 (easy-plane anisotropy), the magnetic hardness parameter loses its meaning~\cite{nieves_database_2019} and is treated as zero in our calculations.

%
\section{Results and discussion}

\subsection{Magnetic properties of CeFe$_{12}$}

\begin{figure}
\vspace{5mm}
\includegraphics[clip,width = 1\columnwidth]{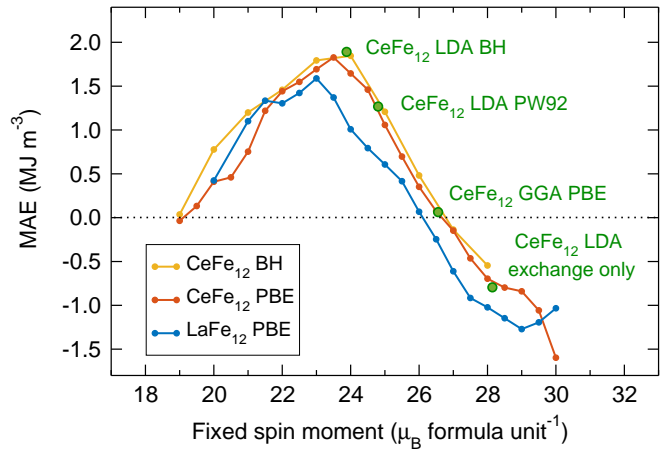}
\caption{\label{fig:CeFe12_MAE_FSM} 
The magnetocrystalline anisotropy energy (MAE) dependence of the fixed spin moment for CeFe$_{12}$ calculated with von Barth-Hedin (BH) and Perdew-Burke-Ernzerhof (PBE) exchange-correlation potentials. Together with corresponding PBE results for LaFe$_{12}$ and equilibrium values for CeFe$_{12}$  obtained for various functionals. The calculations were performed using the FPLO18 code.
}
\end{figure}

Previous calculations for CeFe$_{11}$Ti gave a rather large spread of MAE (from 1.19 to 1.98~MJ\,m$^{-3}$) accompanied by a similarly large scatter of magnetic moments (from 19.19 to 24.04~$\mu_B$\,f.u.$^{-1}$)~\cite{akayama_physical_1994, ke_intrinsic_2016-1, martinez-casado_cefe11ti_2019}.
To confirm the conjecture that the obtained differences have their origin in the choice of exchange-correlation potential,  we performed MAE and magnetic moment calculations for the parent CeFe$_{12}$ phase using the BH, PW92, PBE, and LDA exchange-only functionals. 
The results for various functionals are as follows: BH (MAE~=~1.89~MJ\,m$^{-3}$, $m_S$~=~23.88~$\mu_B$\,f.u.$^{-1}$), PW92 (MAE~=~1.27~MJ\,m$^{-3}$, $m_S$~=~24.80~$\mu_B$\,f.u.$^{-1}$), PBE (MAE~=~0.06~MJ\,m$^{-3}$, $m_S$~=~26.56~$\mu_B$\,f.u.$^{-1}$), and LDA exchange only (MAE~=~-0.79~MJ\,m$^{-3}$, $m_S$~=~28.14~$\mu_B$\,f.u.$^{-1}$). 
As can be seen, the range of MAE values obtained is huge, but interestingly they correlate with the magnetic moment values.
Looking at Fig.~\ref{fig:CeFe12_MAE_FSM}, it is easy to notice that these values (green circles) lie on a common straight line.

Further insight into the observed relationship can be gained by a fully relativistic implementation of the fixed spin moment (FSM) method, which allows one to calculate the MAE as a function of the total spin magnetic moment. 
The results of the MAE($m_S$) calculation for the PBE functional confirm that the spin magnetic moment is an important factor affecting the MAE value. Moreover, the  MAE($m_S$) calculation for another exchange-correlation functional (BH) confirmed that the shape of the MAE($m_S$) function near the equilibrium magnetic moment does not depend on the choice of the functional itself. Hence, we conclude that the MAE($m_S$) function is a more universal fingerprint of the material than the equilibrium MAE values calculated for individual functionals. 

Although different exchange-correlation potentials (BH, PW92, and PBE) lead to different values of equilibrium magnetic moments, for a fixed value of magnetic moment, the band structures determined using the different exchange-correlation potentials are nearly equal, leading to almost identical MAE values and thus to a very similar shape of the MAE($m_S$) function near the equilibrium magnetic moments. Even relatively small changes in the valence band structure have a significant effect on the MAE value \citep{burkert_giant_2004-1, edstrom_magnetic_2015-2}. In the case of induced variations in the spin magnetic moment, the occupancy of the spin channels changes, causing shifting the channels with respect to each other, which further leads to the evolution of the MAE($m_S$) dependence. The inverted-U-like shape of the MAE($m_S$) relation observed for CeFe$_{12}$ in the vicinity of the equilibrium magnetic moment is its individual feature, and the full dependence (from 0) is a function composed of several oscillations, similar to the Fe$_{2}$B case calculated earlier \citep{edstrom_magnetic_2015-2}.

In calculations of the MAE($m_S$) dependence, we observe an almost constant value of the spin magnetic moment on Ce atoms.  Forcing a change in the total spin magnetic moment of the system mainly affects the magnetic moments on the Fe atoms, whose main source is the spin polarization of the 3$d$ orbitals.  Hence, the observed MAE($m_S$) dependence correlates with the evolution of the band structure of the Fe~3$d$ orbitals induced by forcing a change in magnetic moment \citep{edstrom_magnetic_2015-2}. 

The MAE($m_S$) result for LaFe$_{12}$, shown in the Fig.~\ref{fig:CeFe12_MAE_FSM}, will be addressed later in the paper where the effect of 4$f$ electrons on the MAE value will be discussed.
The plots for LaFe$_{12}$ and CeFe$_{12}$ have a similar shape. 
In both cases, the decrease in equilibrium magnetic moment leads to an increase in MAE, which reaches maximum values of 1.58~MJ\,m$^{-3}$ for LaFe$_{12}$ and 1.82~MJ\,m$^{-3}$ for CeFe$_{12}$. 
The spin and orbital magnetic moments of the CeFe$_{12}$ compound are presented in Table~\ref{tab:Mag.Mom}. 

\begin{table}[t]
\caption{\label{tab:Mag.Mom} Spin and orbital magnetic moments ($\mu_B$) on Ce, Fe, and Ti atoms in CeFe$_{11}$Ti (s.g.~{\it Pmmn}, No.~59) and CeFe$_{12}$ (s.g.~$I$4/$mmm$, No.~139). Calculations were performed with FPLO18 using the PBE functional.
}
\begin{tabular}{lll|lll}
 \hline
 \hline
 \multicolumn{3}{c}{CeFe$_{11}$Ti s.g. $Pmmn$} & \multicolumn{3}{c}{CeFe$_{12}$ s.g. $I$4/$mmm$}\\
 \hline
Site &	  ~$m_s$ 	&	   ~~$m_l$     & Site 	&	 ~$m_s$ 	&	   ~$m_l$    \\
 \hline
 Ce & -1.13  & ~ 0.290      &	       Ce &  -1.15 &   0.264     \\
 Fe & ~1.70  & ~ 0.036     	   &	   Fe &  ~2.57  &  0.051 \\
 Fe &  ~1.86 & ~ 0.039         &       Fe &  ~2.38 &   0.047	 \\
 Fe &  ~2.47 & ~ 0.059    	   &	   Fe &  ~1.96 &   0.035  \\
 Fe &  ~2.23 & ~ 0.058	       &        &        &          \\
 Fe &  ~2.08 & ~ 0.050          &	     &        &         \\	
 Fe &  ~2.33 & ~ 0.049          &	     &        &         \\	
 Fe &  ~2.24 & ~ 0.046          &	     &        &         \\	
 Ti & -1.26  & ~ 0.015	        &	     &        &         \\	

\hline 
\hline
\end{tabular}
\end{table}

\subsection{Structural and magnetic properties of CeFe$_{12}$-based compounds with Ti}
\begin{center}
\begin{figure*}
      \centering
 \hfill		\includegraphics[clip,height=0.6\textwidth]{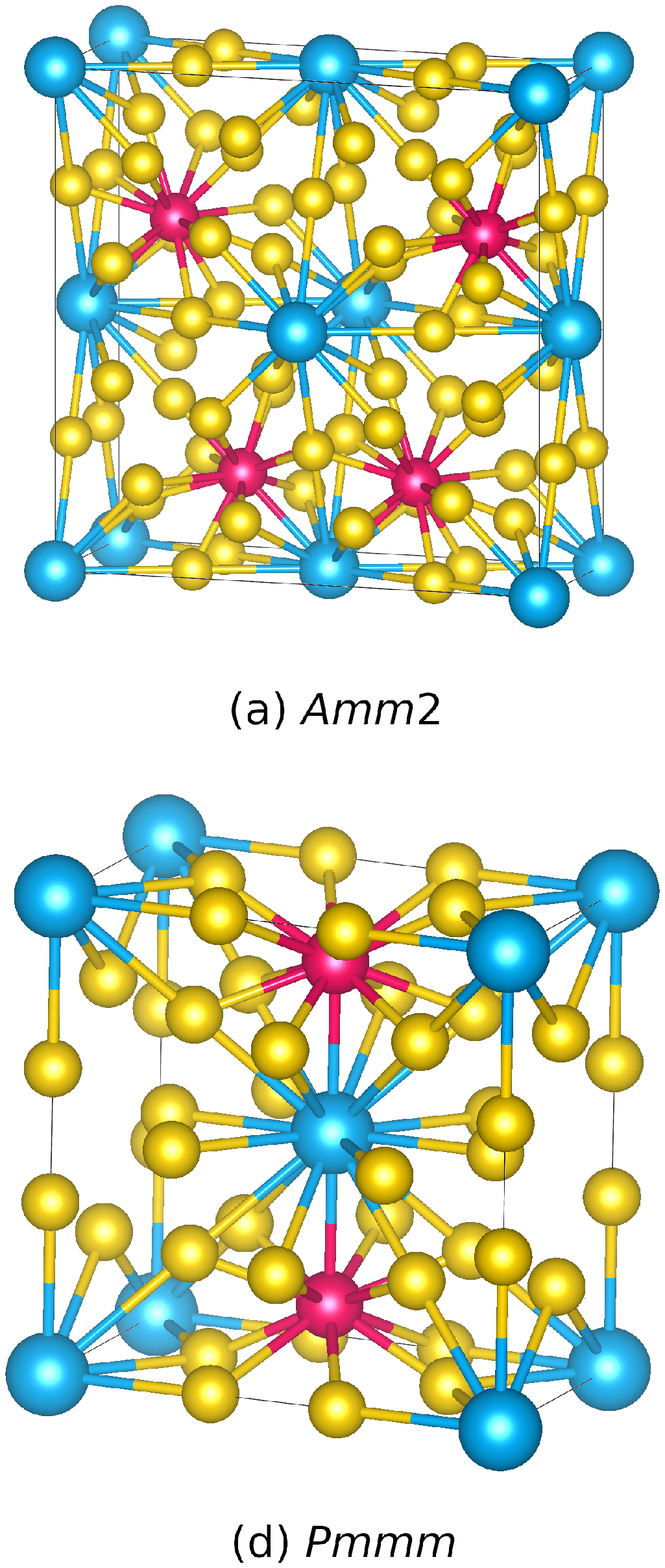}\hfill
       \includegraphics[clip,height=0.6\textwidth]{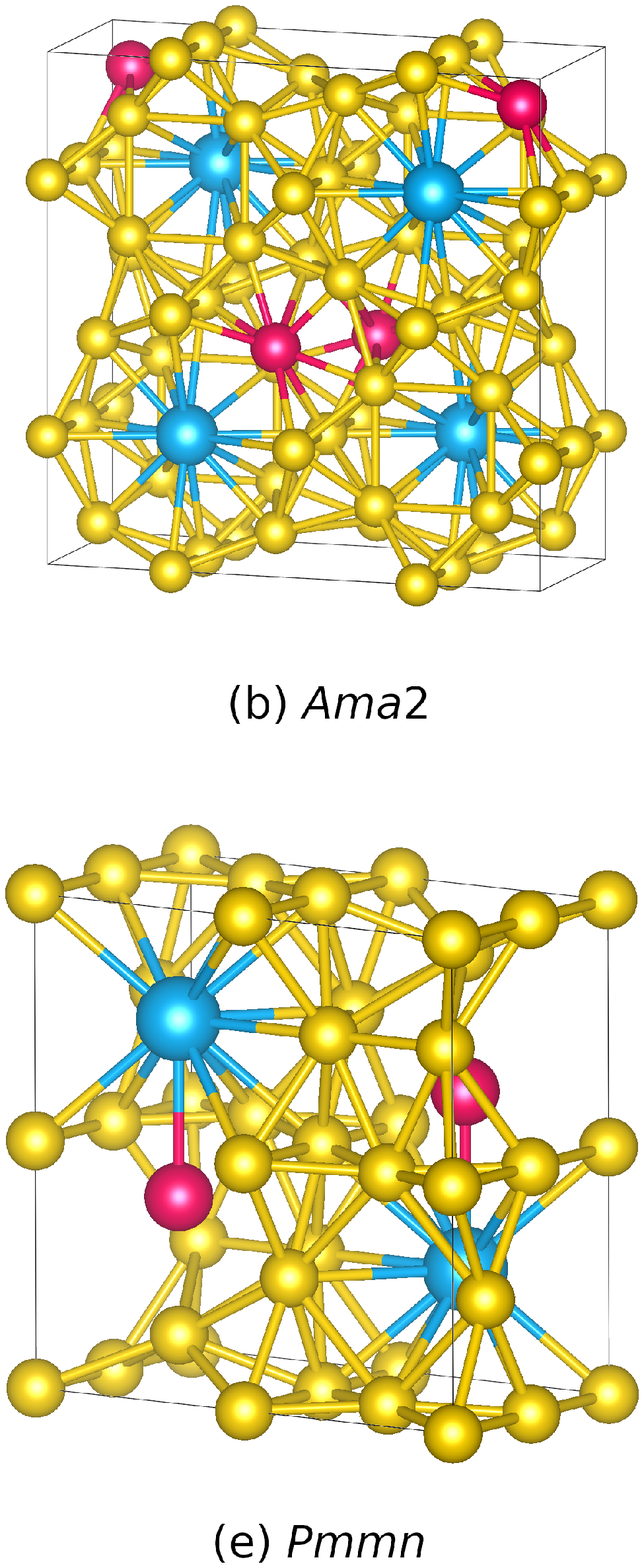}\hfill
       \includegraphics[clip,height=0.6\textwidth]{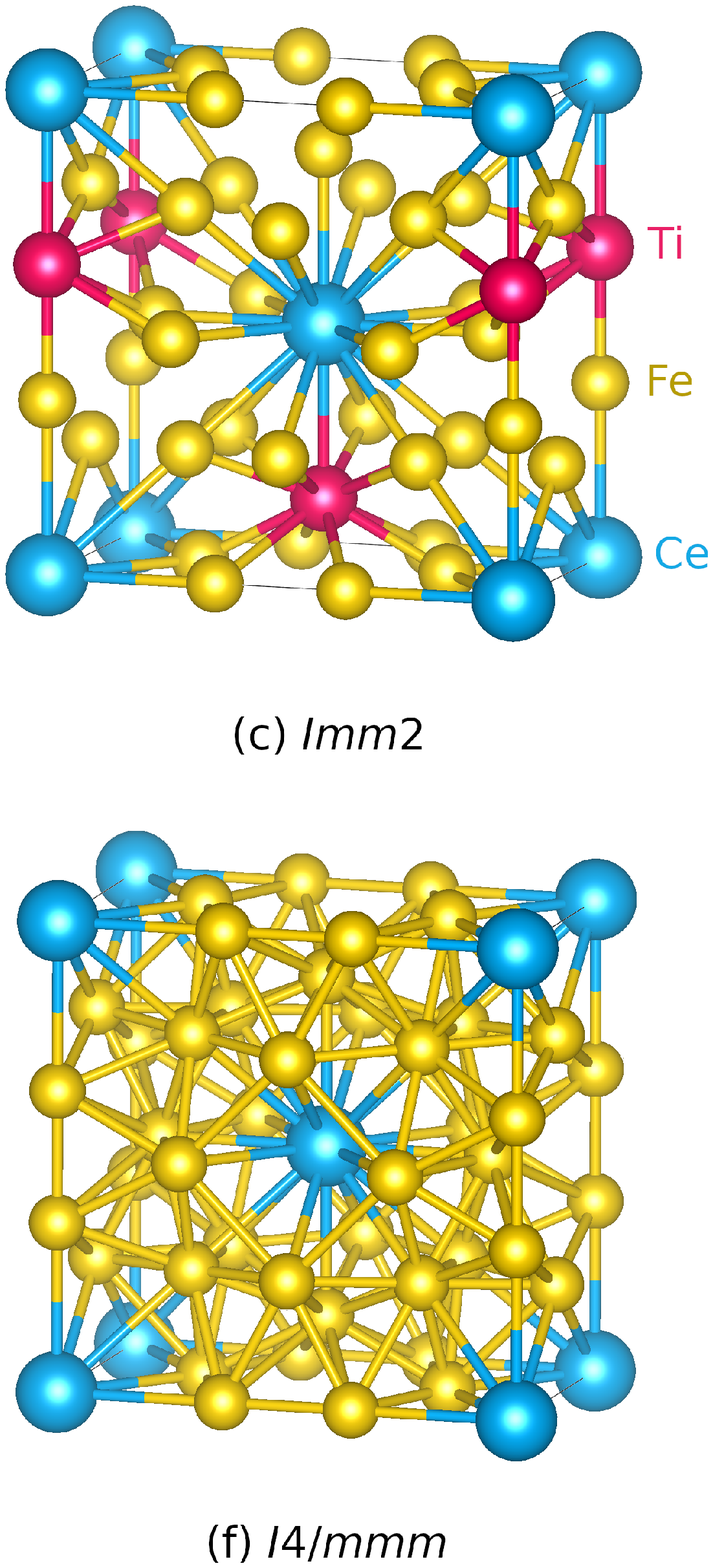} \hfill \hspace{0mm}           
      \caption{
(a--e) Five orthorhombic crystal structures prepared for CeFe$_{11}$Ti compound and (f) tetragonal crystal structure for CeFe$_{12}$.
      \label{fig:fig1}
}
\end{figure*} 
\end{center}
\subsubsection{Supercell model of CeFe$_{11}$Ti}

To find the simplest supercell to model the composition of CeFe$_{11}$Ti, we consider five possible non-equivalent substitution configurations of the dopant atom at the Fe~(8$i$) position, the MAEs, total and partial magnetic moments, and the energy difference relative to the lowest energy structure ($E$~-~$E_0$) for these five configurations are presented in Table~\ref{tab:Struct_type}. 
The cell with space group $Pmmn$ has the lowest energy among the considered structures, so we choose it as the basic unit cell for the calculation of CeFe$_{11}$X compounds presented later in this paper. 
We identify the use of structures with the $Imm$2 space group in earlier works~\cite{delange_crystal-field_2017, galler_intrinsically_2020, martinez-casado_cefe11ti_2019, sozen_ab_2019} as one of the factors that may influence the differences between our and previous results.
MAE range obtained for structures with different space groups is from 1.00 to 1.15~MJ\,m$^{-3}$, and the corresponding range for the total magnetic moment ($m$) is from 21.24 to 22.50~$\mu_B$\,f.u.$^{-1}$.
In order to compare the obtained values for CeFe$_{11}$Ti, the experimental results are presented in Table~\ref{tab:Experimental_CeFe11Ti}. We find that the presented calculated MAEs are in relatively good agreement with the experimentally obtained low-temperature value (1.78~MJ\,m$^{-3}$~\cite{pan_structural_1994}).
\begin{table}[t]
\caption{\label{tab:Struct_type} 
The energy difference between the considered structure and the lowest energy structure $Pmmn$ [$E - E_0$ (meV)], the magnetocrystalline anisotropy energy [MAE~(MJ\,m$^{-3}$)], the spin magnetic moments on Ce [$m_s$(Ce)] and Ti [$m_s$(Ti)], orbital magnetic moment on Ti [$m_l$(Ti)], total magnetic moment [$m$], and magnetic hardness [$\kappa$] calculated for different possible types of structures of CeFe$_{11}$Ti compound. 
Magnetic moments are expressed in $\mu_B$ per atom of formula unit.
Calculations were performed with FPLO18 using the PBE functional. 
}
\begin{tabular}{p{1.2cm} S[table-format=3.2]  S[table-format=1.2]  S[table-format=1.2]  S[table-format=1.2]  S[table-format=1.3]  S[table-format=2.2] S[table-format=1.2]}
 \hline
 \hline
S.g.	&	{$E$-$E_0$}	&	{MAE}	&	{$m_s$(Ce)} &	{$m_s$(Ti)}	&	{$m_l$(Ti)}	&	{$m$} & {$\kappa$}	\\
\hline											
$Amm$2	&	31.2	&	1.14	&	-1.04  &	-1.24	&	0.014 	&	22.01  &   0.81  \\
$Ama$2	&	116.8	&	1.08	&	-1.01  &	-1.22	&	0.014 	&	21.99  &   0.79  \\
$Imm$2	&	39.8	&	1.15	&	-1.10   & 	-1.26	&	0.014 	&	21.92  &   0.82  \\
$Pmmm$	&	344.6	&	1.02	&	-1.08  & 	-1.26	&	0.010 	&	22.50   &   0.75  \\
$Pmmn$	&	0.0	    &	1.00	&	-1.10  & 	-1.30	&	0.015 	&	21.24   &   0.79  \\
\hline
\hline
\end{tabular}
\end{table}
\begin{figure}
\vspace{5mm}
\includegraphics[clip,width = 0.95\columnwidth]{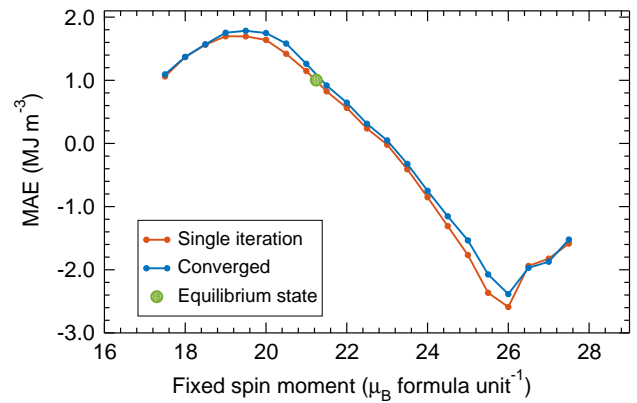}
\caption{\label{fig:CeFe11Ti_MAE_1itvsCONV} 
Magnetocrystalline anisotropy energies (MAEs) as a function of fixed spin moment for the CeFe$_{11}$Ti compound calculated using full-convergence method (blue line) and the method using a single fully relativistic iteration (orange line). The green point represents the equilibrium value. The calculations were performed using FPLO18 with the PBE functional.
}
\end{figure}

\begin{table}
\caption{\label{tab:Experimental_CeFe11Ti} Experimental values of magnetocrystalline anisotropy energy [MAE~(MJ\,m$^{-3}$)] for CeFe$_{11}$Ti compound obtained from magnetization curves. }
\begin{tabular}{p{1cm}p{1.0cm}p{1.0cm}}
 \hline
 \hline
Ref.	&	MAE	& $T$(K)	\\
\hline											
\cite{pan_structural_1994}	&	0.62	& 300	\\
\cite{akayama_physical_1994}	    &	1.10 & 300	\\
\cite{pan_structural_1994}	&	0.76 & 300	\\
\cite{pan_structural_1994} & 1.78 & 1.5 \\
\hline
\hline
\end{tabular}
\end{table}

We can see in Table~\ref{tab:Struct_type} that the spin magnetic moment on the Ce atom is close to -1~$\mu_B$ in all cases. The negative sign in this notation indicates the magnetic moment direction opposite to the ferromagnetic Fe matrix. In the case of the $Pmmn$ structure, the spin and orbital magnetic moments on the Ce, Fe, and Ti are shown in Table~\ref{tab:Mag.Mom}.
Comparing the CeFe$_{12}$ and the Ti doped counterpart, differences in both spin and orbital moments can be observed. 
For CeFe$_{12}$, the spin magnetic moment on the Fe atom is in the range of 1.96-2.57~$\mu_B$, and the corresponding orbital magnetic moment is 0.035-0.051~$\mu_B$. In the case of CeFe$_{11}$Ti, the spin magnetic moment on the Fe atom is in the range of 1.7-2.47~$\mu_B$ and the orbital magnetic moment is 0.036-0.059~$\mu_B$. For both systems, the spin magnetic moment on the Ce atom is similar, however, for the Ti-doped system, the orbital magnetic moment on the Ce atom decreases by 0.026~$\mu_B$.

\subsubsection{Effect of Ti concentration on magnetic properties}
\begin{table}
\caption{\label{tab:Substit_12-11} 
Magnetocrystalline anisotropy energy  [MAE~(MJ\,m$^{-3}$)], total magnetic moment [$m$~($\mu_B$\,f.u.$^{-1}$)], and magnetic hardness [$\kappa$] for selected CeFe$_{12-y}$Ti$_{y}$ compounds. The calculations were performed with FPLO18 using the PBE functional.
}
\begin{tabular}{p{2cm}p{1.5cm}p{1cm}p{1cm}p{1.0cm}}
 \hline
 \hline
Compound	              & S.g. & MAE   &	$m$  & $\kappa$	\\
\hline											
CeFe$_{12}$               & $I$4/$mmm$   &  0.06   &  26.55     &  0.16    \\
CeFe$_{11.5}$Ti$_{0.5}$   & $Pmm$2       &  0.59   &  24.68     &  0.52   \\
CeFe$_{11}$Ti             & $Pmmn$       &  1.00   &  21.24     &  0.79     \\
CeFe$_{10}$Ti$_{2}$	      & $P$4/$mmm$   &	1.30   &  18.04	    &  1.06  \\
\hline
\hline
\end{tabular}
\end{table} 

Comparing the presented computational results for CeFe$_{12}$ and CeFe$_{11}$Ti, we expect that a further increase in the amount of substituent should lead to a further decrease in the total magnetic moment, which in turn, due to the MAE($m_S$) relation, should lead to an increase in the MAE value. This prediction is tentatively confirmed by the computational results for CeFe$_{12-y}$Ti$_{y}$, where $y$~=~0,~0.5,~1,~2, presented in Table~\ref{tab:Substit_12-11}. We see that the magnetic moment strongly decreases with increasing concentration of Ti and observe  a close to linear dependence of MAE($y$). We also see that for $y$~=~2 the result of magnetic hardness is clearly above 1, making this material a potential new permanent magnet.

\subsubsection{Fixed spin moment calculations}

Figure~\ref{fig:CeFe11Ti_MAE_1itvsCONV} presenting the MAE($m_S$) relation for CeFe$_{11}$Ti is very similar to the analogous result for CeFe$_{12}$ shown earlier, see Fig.~\ref{fig:CeFe12_MAE_FSM}.
In both cases, we observe a parabola-shaped relationship with a maximum below the equilibrium magnetic moment. 
For CeFe$_{11}$Ti, the MAE maximum of 1.78~MJ\,m$^{-3}$ occurs for a total spin magnetic moment of 19.5~$\mu_B$\,f.u.$^{-1}$, while the equilibrium MAE value is 1.00~MJ\,m$^{-3}$ for $m_S$~=~21.2~$\mu_B$\,f.u.$^{-1}$.
Comparing again the results for CeFe$_{11}$Ti and CeFe$_{12}$, we can interpret the increase in the equilibrium MAE value for CeFe$_{11}$Ti as the effect of the doping-induced decrease in the total magnetic moment of the system. Furthermore, a comparison of calculations based on a single fully relativistic iteration shown in Fig.~\ref{fig:CeFe11Ti_MAE_1itvsCONV} with the fully convergent calculations positively verifies the approximation applied, the main advantage of which is a multiple reduction in computation time.

\subsubsection{Effect of interstitial dopants on magnetic hardness}

\begin{table}[h!]
\caption{\label{tab:Substit_CeFe12X} 
The magnetocrystalline anisotropy energy [MAE~ (MJ\,m$^{-3}$)], 
total magnetic moment [$m$~($\mu_B$\,f.u.$^{-1}$)], and 
magnetic hardness [$\kappa$] for CeFe$_{12}$-based and CeFe$_{11}$Ti-based compounds with interstitial dopants H, B, C, and N. 
The calculations were performed with the FPLO18 code using the PBE functional.
}
\begin{tabular}{p{2cm}p{1cm}p{1cm}p{1.0cm}}
 \hline
 \hline
Compound	               & MAE   &	$m$  & $\kappa$	\\
\hline					
CeFe$_{12}$                &  0.06   &  26.55     &  0.16    \\	
CeFe$_{12}$H               &  1.12 &  25.83     &  0.69   \\             	
CeFe$_{12}$C               &  1.50  &  25.39     &  0.81    \\
CeFe$_{12}$N               &  1.82 &  27.03     &  0.84     \\
CeFe$_{11}$Ti             &  1.00   &  21.24     &  0.79     \\
CeFe$_{11}$TiB               &  1.52  &  20.96     &  0.99    \\
CeFe$_{11}$TiN              &  1.64 &  22.53     &  0.95   \\
\hline
\hline
\end{tabular}
\end{table} 
To shift the position on the MAE($m_S$) curve from the equilibrium location towards the observed maximum, we consider the addition of  light atoms such as carbon, boron, nitrogen, or hydrogen at the interstitial positions of CeFe$_{12}$ and CeFe$_{11}$Ti. From the results presented in Table~\ref{tab:Substit_CeFe12X}, it can be seen that nitridation allows us to approach the maximum MAE value for the pure CeFe$_{12}$ compound.
The results for CeFe$_{11}$TiB  and CeFe$_{11}$TiN show that doping with a light atom can increase both the MAE value and the magnetic hardness of the resulting compound.

\subsection{Intrinsic magnetic properties of \\CeFe$_{11}$X and CeFe$_{10}$X$_{2}$ compounds}

\subsubsection{Magnetocrystalline anisotropy energy}
\begin{figure*}
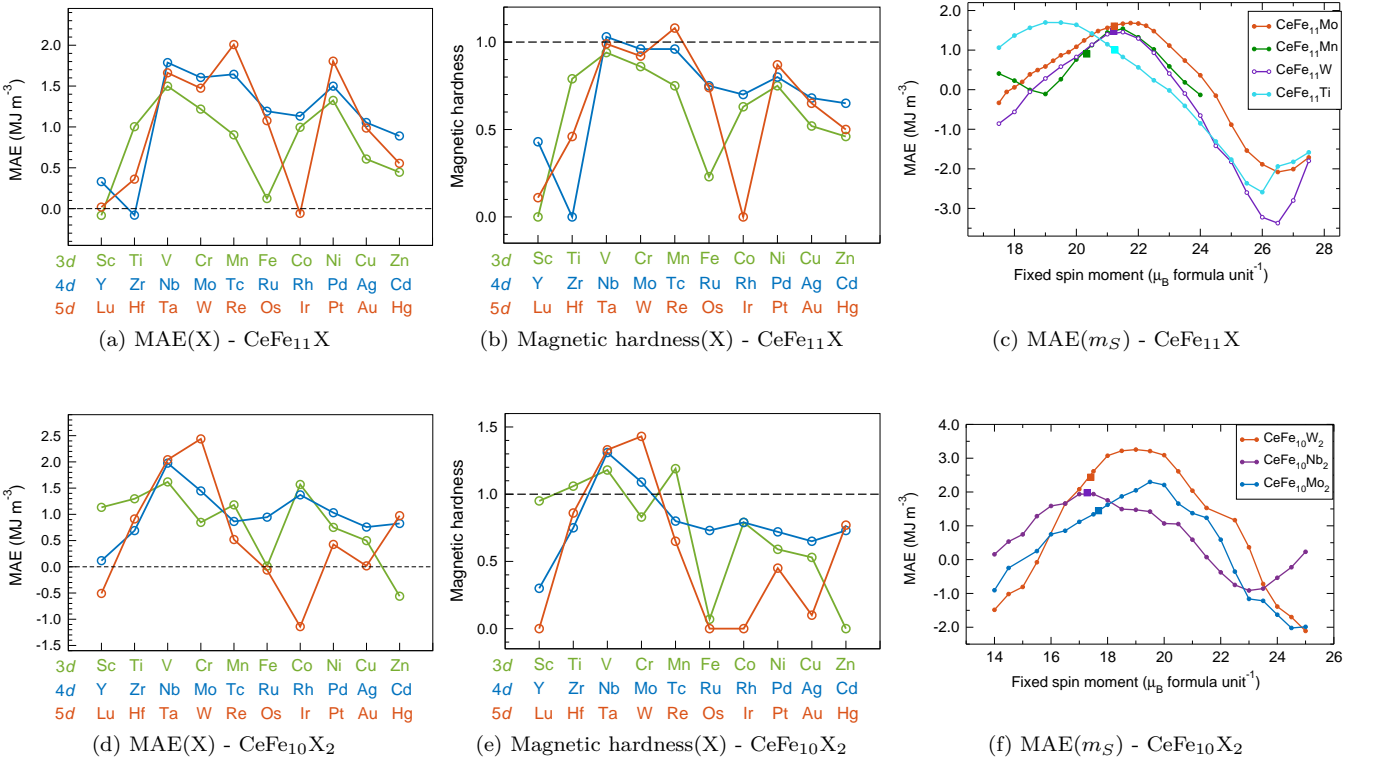

\centering
\subfloat[MAE(X) - CeFe$_{11}$X]{\label{MAE_"1-11-1"}
\includegraphics[clip, width=0.32\textwidth]{CeFe11.MAE_3d_4d_5d.eps}}
\hfill
\subfloat[Magnetic hardness(X) - CeFe$_{11}$X]{\label{kappa_"1-11-1"}
\includegraphics[clip, width=0.32\textwidth]{CeFe11.Mag_Hardness_3d_4d_5d.eps}}
\hfill
\subfloat[MAE($m_S$) - CeFe$_{11}$X]{\label{FSM_"1-11-1"}
\includegraphics[clip, width=0.33\textwidth]{cefe11X_mae_vs_m.eps}}
\vspace{4mm}
\subfloat[MAE(X) - CeFe$_{10}$X$_{2}$]{\label{MAE_"1-10-2"}
\includegraphics[clip, width=0.32\textwidth]{CeFe10X2_MAE_3d_4d_5d.eps}}
\hfill
\subfloat[Magnetic hardness(X) - CeFe$_{10}$X$_{2}$]{\label{kappa_"1-10-2"}
\includegraphics[clip, width=0.32\textwidth]{CeFe10X2_Mag_Hardness_3d_4d_5d.eps}}
\hfill
\subfloat[MAE($m_S$) - CeFe$_{10}$X$_{2}$]{\label{FSM_"1-10-2"}
\includegraphics[clip, width=0.33\textwidth]{cefe10W2_FSM.eps}}
\caption{
The hard-magnetic properties of CeFe$_{11}$X and CeFe$_{10}$X$_{2}$ systems calculated for various 3$d$, 4$d$, and 5$d$ transition metals.
(a, d)~the magnetocrystalline anisotropy energy, 
(b, e)~the magnetic hardness, and
(c, f)~magnetocrystalline anisotropy energy dependencies of fixed spin moment for the selected elements (X~=~Mo, Mn, W, Ti for CeFe$_{11}$X and X~=~W, Nb, Mo for CeFe$_{10}$X$_2$). 
The squares in the MAE($m_S$) plots represent equilibrium values.
The calculations were performed with FPLO18 using the PBE functional and supercell approach. 
}
\end{figure*}

Since we considered the doping of CeFe$_{11}$X and CeFe$_{10}$X$_{2}$ with all 3$d$, 4$d$, and 5$d$ transition metals, we obtained a complete picture of the MAE changes depending on the type and amount of substituent, see Figs.~\ref{MAE_"1-11-1"}~and~\ref{MAE_"1-10-2"} and Table~\ref{tab:MAE- CeFe11X|CeFe10X2}.
Moreover, the dependence of magnetic hardness ($\kappa$) on transition metal is shown in Figs.~\ref{kappa_"1-11-1"} and \ref{kappa_"1-10-2"}.
It can be seen that the magnetic hardness results are very similar to the MAE trends. 
Among all CeFe$_{11}$X compositions, only CeFe$_{11}$Nb and CeFe$_{11}$Re can be classified as hard permanent magnets ($\kappa$ > 1), while the values of $\kappa$ for  CeFe$_{11}$W and CeFe$_{11}$Mo are slightly below the classification criterion, which is still a good indication for further modifications. In the context of potential applications, in addition to the physical parameters, the economic aspect must be taken into account.
Thus, although we observe that also Au and Pt significantly increase magnetic hardness, it is difficult to imagine their practical applications. 
Considering the raw material prices, the economically justified dopant group will include Ti, Cr, Ni, W, Mo, Mn, and Nb. 
The MAE($m_S$) dependencies for the selected Ti, W, Mo, and Mn dopants in"1-11-1" system are shown in Fig.~\ref{FSM_"1-11-1"}. It can be seen that in each of these cases the CeFe$_{11}$X compound has the potential to reach the MAE of about 1.5~MJ\,m$^{-3}$ and $\kappa$~>~1. 

MAE calculations of the CeFe$_{10}$X$_{2}$ compositions show that: CeFe$_{10}$W$_{2}$, CeFe$_{10}$Nb$_{2}$, CeFe$_{10}$Mo$_{2}$, and CeFe$_{10}$Mn$_{2}$ can be classified as hard permanent magnets and can be worthy of further investigations.     
We also calculated MAE($m_S$) curves for the selected compounds to show that, especially in the case of the CeFe$_{10}$W$_{2}$ compound, there is a possible range for further modification to obtain very high MAE even above 3~MJ\,m$^{-3}$, see Fig.~\ref{FSM_"1-10-2"}.
From our enthalpy of formation calculations, it is clear that CeFe$_{11}$Ti is chemically stable, whereas the considered compounds with W, Mo and Mn are not. In this situation, to obtain the indicated compounds with promising hard magnetic properties, it may be necessary to stabilize them, for example, by partial replacement of Ce with other rare earth elements or complementary replacement of some Fe atoms with an additional element \citep{martinez_sanchez_effect_2020, dasmahapatra_doping_2021}. 
Zhou and Pinkerton described the magnetic hardening of CeFe$_{12-x}$Mo$_{x}$ by melt spinning, while revealing a complex multiphase microstructure with "1-12" phase contents ranging from 78 to 87 wt\% \citep{zhou_magnetic_2014}.
Our enthalpy of formation calculations suggest chemical instabillity of CeFe$_{11}$Mo and CeFe$_{10}$Mo$_{2}$, which may explain the experimental difficulties in obtaining a homogeneous phase of nominal composition.

\begin{figure}
\vspace{5mm}
\includegraphics[clip,width = 1\columnwidth]{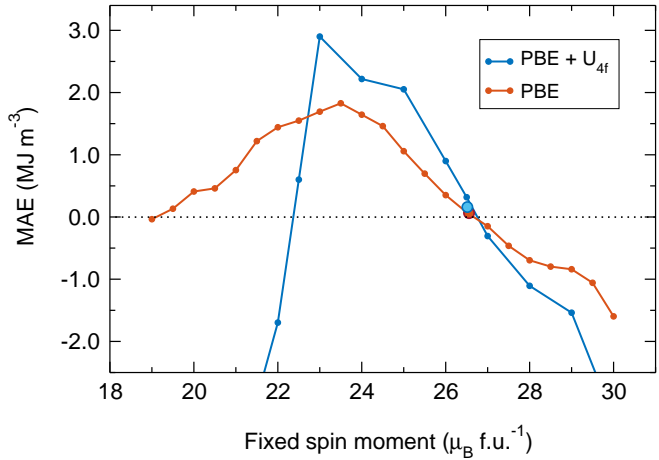}
\caption{\label{fig:PBE+U_FSM} 
The magnetocrystalline anisotropy energy (MAE) dependence of the fixed spin moment for CeFe$_{12}$ calculated with PBE~+~U exchange-correlation potential, for U equal to 0 and 3~eV, together with corresponding equilibrium values denoted by larger filled circles. The calculations were performed using the FPLO18 code.
}
\end{figure}
\begin{figure*}
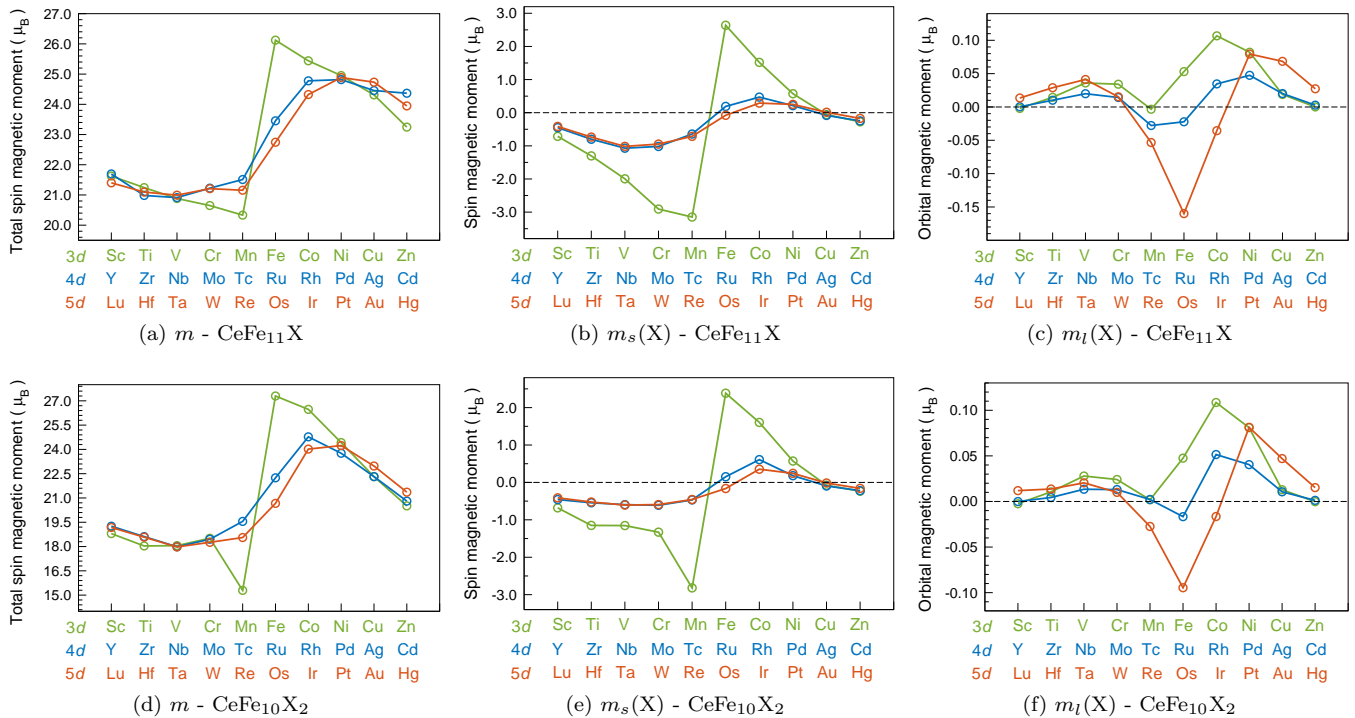

\centering
\subfloat[$m$ - CeFe$_{11}$X]{\label{m_"1-11-1"}
\includegraphics[clip, width=0.32\textwidth]{CeFe11_Total_Spin_Mom_3d_4d_5d.eps}}
\hfill
\subfloat[$m_s$(X) - CeFe$_{11}$X]{\label{ms_"1-11-1"}
\includegraphics[clip, width=0.32\textwidth]{CeFe11.ms_3d_4d_5d.eps}}
\hfill
\subfloat[$m_l$(X) - CeFe$_{11}$X]{\label{ml_"1-11-1"}
\includegraphics[clip, width=0.32\textwidth]{CeFe11.ml_3d_4d_5d.eps}}
\hfill
\subfloat[$m$ - CeFe$_{10}$X$_{2}$]{\label{m_"1-10-2"}
\includegraphics[clip, width=0.32\textwidth]{CeFe10X2_Total_Spin_Mom_3d_4d_5d.eps}}
\hfill
\subfloat[$m_s$(X) - CeFe$_{10}$X$_{2}$]{\label{ms_"1-10-2"}
\includegraphics[clip, width=0.32\textwidth]{CeFe10X2_ms_3d_4d_5d.eps}}
\hfill
\subfloat[$m_l$(X) - CeFe$_{10}$X$_{2}$]{\label{ml_"1-10-2"}
\includegraphics[clip, width=0.32\textwidth]{CeFe10X2_ml_3d_4d_5d.eps}}
\caption{\label{fig:6}
The magnetic moments of CeFe$_{11}$X and CeFe$_{10}$X$_{2}$ systems calculated for various 3$d$, 4$d$, and 5$d$ transition metals X.
(a, d) 
Total magnetic moments per formula unit and
(b, c, e, f) 
spin and orbital magnetic moments on dopant element X.
The calculations were performed with FPLO18 using the PBE functional for quantization axis~[010]. 
}
\end{figure*}

\subsubsection{Magnetic moments}

The results shown in Figures~\ref{ms_"1-11-1"} and \ref{ms_"1-10-2"} indicate that none of the considered dopant elements at position 8$i$ has a higher spin magnetic moment than the corresponding Fe atom in the reference compound CeFe$_{12}$. For the CeFe$_{11}$X compounds considered, the range of spin magnetic moments on the Ce atom is from -0.88 to -1.15~$\mu_B$, and the orbital magnetic moments on the Ce atom is from 0.14 to 0.31~$\mu_B$. For doping with Mn, we observe the lowest values of both the total spin magnetic moment and the partial magnetic moment. We also observe that the function of the spin magnetic moments on the substitutions resembles a sinusoidal function with a minimum and maximum on the positive and negative side of the value interval, see Fig.~\ref{fig:6}. 
Results obtained are consistent with previous calculations of the dependence of the total spin magnetic moment on the position of substitution, for example of the 3$d$ elements in NdFe$_{11}$X system \cite{harashima_first-principles_2016}. 
Analogous trends in the substitution of 5$d$ elements in the 3$d$ matrix have been previously found computationally \cite{akai_nuclear_nodate, dederichs_ab-initio_1991-1, werwinski_magnetocrystalline_2018} and experimentally \cite{wienke_determination_1991-1}.

\subsubsection{Effect of 4$f$ electrons on magnetocrystalline anisotropy}

In the heavy-fermion regime, the local magnetic moment of Ce is shielded by conduction electrons that couple to the single Ce 4$f$ electron to form a nonmagnetic single state~\cite{galler_intrinsically_2020}. This kind of heavy-fermionic behavior of Ce has been observed in families "1-5" (CeCo$_5$)\cite{bartashevich_magnetic_1994} and "2-14-1" (Ce$_2$Fe$_{14}$B)~\cite{capehart_steric_1993} of hard-magnetic intermetallics. 
To determine the effect of 4$f$ electrons on the obtained MAE values, we performed MAE($m_S$) calculations for an isostructural LaFe$_{12}$ compound with an empty 4$f$ shell, see Fig.~\ref{fig:CeFe12_MAE_FSM}. 
The relatively small differences observed between the MAE($m_S$) curves calculated for the LaFe$_{12}$ and CeFe$_{12}$ compounds indicate that, at least in the PBE approximation, the effect of the 4$f$ shell on the MAE values is small.
Subsequent MAE  calculations for  LaFe$_{11}$Ti (0.66~MJ\,m$^{-3}$), CeFe$_{11}$Ti (1.00~MJ\,m$^{-3}$), LaFe$_{10}$W$_{2}$ (1.91~MJ\,m$^{-3}$), and CeFe$_{10}$W$_{2}$ (2.43~MJ\,m$^{-3}$) also confirm the secondary effect of the 4$f$ shell on the MAE values in these compounds. 
Since La and Ce are extracted in large quantities during the mining of Nd-containing rare earth ores, both of these elements may play an important role in the development of lower-cost high-performance permanent magnets.

Finally, we would like to briefly discuss how the use of the intra-atomic Hubbard repulsion term U (GGA~+~U) affects the obtained results. For this purpose, we performed PBE~+~U calculations of the magnetic properties for the example compound CeFe$_{12}$. With increasing U in the range from 0 to 3~eV, the equilibrium total spin magnetic moment slightly decreases, see Table~\ref{tab:PBE+U}. The introduction of intra-atomic repulsion also affects the MAE value raising it from 0.06 MJ\,m$^{-3}$  for U~=~0 to 0.16 MJ\,m$^{-3}$ for U~=~3~eV. The observed changes are rather not qualitative, which allow us to conclude that the use of the PBE approximation in this work was justified. The results of calculations of the magnetocrystalline anisotropy energy (MAE) dependence on the fixed spin moment with PBE~+~U, see Fig.~\ref{fig:PBE+U_FSM}, lead to similar conclusions. Although the differences between the results without and with Hubbard U correction are clear, in both cases we obtain similar equilibrium state properties and qualitatively similar relations showing the MAE maximum for reduced magnetic moments.
\begin{table}
\caption{\label{tab:PBE+U} 
Total spin magnetic moments [$m_{S}$ ($\mu_B$\,f.u.$^{-1}$)] and magnetocrystalline anisotropy energies [MAE (MJ\,m$^{-3}$)] calculated for CeFe$_{12}$ with different values of Hubbard U (in eV) for PBE~+~U approach, with U applied to the Ce 4$f$ orbitals. The calculations were performed using the FPLO18 code.
}
\begin{tabular}{p{1cm}p{1cm}p{1.0cm}}
 \hline
 \hline
U	&	MAE 	&	 $m_{S}$ 	\\
\hline											
0	&	0.06	&	26.56	\\
1	&	0.25	&	26.55	\\
2	&	0.11	&	26.54	\\
3	&	0.16	&	26.52	\\
\hline
\hline
\end{tabular}
\end{table}

\section{Summary and conclusions}
Using density functional theory, we investigated the magnetic properties of the compounds CeFe$_{11}$X and CeFe$_{10}$X$_2$ with all 3$d$, 4$d$, and 5$d$ transition metals. 
%
%
As a starting example we used the 
CeFe$_{11}$Ti system, for which we determined the simplest model of the crystal structure and further studied how changing the Ti concentration affects the obtained magnetic properties.
We found that in the doping range of 0 to 2 Ti atoms per 12 Fe atoms, an increase in Ti concentration leads to a decrease in the total magnetic moment and a significant increase in the magnetocrystalline anisotropy energy (MAE) and magnetic hardness.
Using the fully relativistic fixed spin moment method, we calculated for CeFe$_{11}$Ti the dependence of the MAE on the spin magnetic moment, which confirmed the expected relation that MAE increases with decreasing magnetic moment.

%
An almost identical dependence of the magnetocrystalline anisotropy energy on the magnetic moment as for CeFe$_{11}$Ti was observed for the parent compound CeFe$_{12}$.
We also showed that the discussed relation of the anisotropy energy on the magnetic moment practically does not depend on the choice of the exchange-correlation potential.
On the other hand, the equilibrium magnetic moment values obtained depend very strongly on the choice of the exchange-correlation potential, which indirectly affects the anisotropy energy values and leads to a very wide range of results, from strongly negative to strongly positive.
Additional calculations for CeFe$_{12}$- and CeFe$_{11}$Ti-based alloys with smaller elements such as B, C, and N placed in interstitial positions showed that such doping can lead to a decrease in the total magnetic moment and an accompanying increase in the MAE.

%
After a detailed analysis of the structure model of CeFe$_{11}$Ti and verification of the fixed spin moment method, we performed systematic calculations for the CeFe$_{11}$X and CeFe$_{10}$X$_2$ compounds considering the full range of transition metal dopants.
We have presented and discussed the courses of dependence of magnetic moments and magnetocrystalline anisotropy energy on the type of dopant.
Compositions showing very high magnetic hardness include CeFe$_{11}$W, CeFe$_{10}$W$_{2}$, CeFe$_{11}$Mn, CeFe$_{10}$Mn$_{2}$, CeFe$_{11}$Mo, CeFe$_{10}$Mo$_{2}$, and CeFe$_{10}$Nb$_{2}$.

%
Furthermore, we  also determined the effect of 4$f$ electrons on the obtained MAE values by performing MAE($m_S$) calculations for isostructural LaFe$_{12}$ and equilibrium calculations of LaFe$_{11}$Ti and LaFe$_{10}$W$_{2}$ compounds with empty 4$f$ shell. 
By comparing the obtained results for the corresponding systems with Ce, we found a secondary effect of the 4$f$ shell on the MAE value in these systems.
We also found that the application of the intra-atomic Hubbard repulsion term U (GGA~+~U) for the CeFe$_{12}$ test case does not induce qualitative differences in the obtained results between GGA and GGA~+~U results, which justifies the use of the PBE approximation throughout the paper.

\begin{table*}
\caption{\label{tab:MAE- CeFe11X|CeFe10X2} 
The magnetocrystalline anisotropy energy [MAE (MJ\,m$^{-3}$)], 
total magnetic moment [$m$], 
magnetic hardness [$\kappa$],
orbital magnetic moment [$m_l$(X)], and
spin magnetic moment [$m_s$(X)]
calculated for CeFe$_{11}$X (s.g.~{\it Pmmn}, No.~59) and CeFe$_{10}$X$_{2}$ (s.g.~$P$4/$mmm$, No.~123) with various 3$d$, 4$d$, and 5$d$ transition metal elements X in compounds. 
Magnetic moments are expressed in $\mu_B$ per atom of formula unit.
Calculations were performed with FPLO18 using the PBE functional. 
}
\begin{ruledtabular}
\begin{tabular}{lrrrrrlrrrrr}

\multicolumn{6}{c}{CeFe$_{11}$X} & \multicolumn{6}{c}{CeFe$_{10}$X$_{2}$}		\\	
\hline
\multicolumn{12}{c}{3$d$ elements}		\\																
\hline																								
    &              MAE      &       $m$   &       $\kappa$        &       $m_l$(X)   &       $m_s$(X)	&	        &              MAE      &       $m$ &       $\kappa$        &       $m_l$(X)   &       $m_s$(X)	\\																				
																							
\hline																							
CeFe$_{11}$Sc	&	-0.08	&	22.37	&	0.00	&	-0.002	&	-0.71	&	CeFe$_{10}$Sc$_{2}$	&	1.13	&	18.80	&	0.95	&	-0.003	&	-0.68	\\
CeFe$_{11}$Ti	&	1.00	&	21.24	&	0.79	&	0.015	&	-1.30	&	CeFe$_{10}$Ti$_{2}$	&	1.30	&	18.04	&	1.06	&	0.011	&	-1.15	\\
CeFe$_{11}$V	&	1.50	&	21.75	&	0.94	&	0.036	&	-2.00	&	CeFe$_{10}$V$_{2}$	&	1.62	&	18.06	&	1.18	&	0.028	&	-1.15	\\
CeFe$_{11}$Cr	&	1.22	&	21.51	&	0.86	&	0.034	&	-2.91	&	CeFe$_{10}$Cr$_{2}$	&	0.85	&	18.51	&	0.83	&	0.024	&	-1.33	\\
CeFe$_{11}$Mn	&	0.90	&	21.14	&	0.75	&	-0.003	&	-3.15	&	CeFe$_{10}$Mn$_{2}$	&	1.18	&	15.29	&	1.19	&	0.002	&	-2.82	\\
CeFe$_{11}$Fe	&	0.12	&	26.12	&	0.23	&	0.053	&	2.64	&	CeFe$_{10}$Fe$_{2}$	&	0.01	&	27.30	&	0.07	&	0.048	&	2.38	\\
CeFe$_{11}$Co	&	1.00	&	26.38	&	0.63	&	0.107	&	1.52	&	CeFe$_{10}$Co$_{2}$	&	1.57	&	26.47	&	0.79	&	0.109	&	1.60	\\
CeFe$_{11}$Ni	&	1.32	&	25.87	&	0.75	&	0.082	&	0.57	&	CeFe$_{10}$Ni$_{2}$	&	0.75	&	24.42	&	0.59	&	0.081	&	0.57	\\
CeFe$_{11}$Cu	&	0.61	&	25.15	&	0.52	&	0.019	&	-0.06	&	CeFe$_{10}$Cu$_{2}$	&	0.50	&	22.32	&	0.53	&	0.013	&	-0.08	\\
CeFe$_{11}$Zn	&	0.45	&	24.07	&	0.46	&	0.000	&	-0.27	&	CeFe$_{10}$Zn$_{2}$	&	-0.56	&	20.51	&	0.00	&	0.000	&	-0.24	\\
\hline																							
\multicolumn{12}{c}{4$d$ elements}		\\																				
\hline															
        &              MAE      &       $m$   &       $\kappa$        &       $m_l$(X)   &       $m_s$(X)	&	        &              MAE      &       $m$ &       $\kappa$        &       $m_l$(X)   &       $m_s$(X)	\\																				
																							
\hline																							
CeFe$_{11}$Y	&	0.33	&	22.39	&	0.43	&	0.000	&	-0.45	&	CeFe$_{10}$Y$_{2}$	&	0.12	&	19.26	&	0.30	&	0.000	&	-0.46	\\
CeFe$_{11}$Zr	&	-0.08	&	21.72	&	0.00	&	0.010	&	-0.80	&	CeFe$_{10}$Zr$_{2}$	&	0.69	&	18.62	&	0.75	&	0.004	&	-0.54	\\
CeFe$_{11}$Nb	&	1.79	&	21.77	&	1.03	&	0.020	&	-1.07	&	CeFe$_{10}$Nb$_{2}$	&	1.98	&	17.98	&	1.31	&	0.013	&	-0.60	\\
CeFe$_{11}$Mo	&	1.61	&	22.06	&	0.96	&	0.014	&	-1.02	&	CeFe$_{10}$Mo$_{2}$	&	1.45	&	18.45	&	1.09	&	0.013	&	-0.61	\\
CeFe$_{11}$Tc	&	1.64	&	22.30	&	0.96	&	-0.028	&	-0.64	&	CeFe$_{10}$Tc$_{2}$	&	0.86	&	19.55	&	0.80	&	0.002	&	-0.47	\\
CeFe$_{11}$Ru	&	1.19	&	24.26	&	0.75	&	-0.022	&	0.19	&	CeFe$_{10}$Ru$_{2}$	&	0.95	&	22.25	&	0.73	&	-0.017	&	0.15	\\
CeFe$_{11}$Rh	&	1.13	&	25.65	&	0.70	&	0.035	&	0.47	&	CeFe$_{10}$Rh$_{2}$	&	1.37	&	24.78	&	0.79	&	0.052	&	0.61	\\
CeFe$_{11}$Pd	&	1.50	&	25.71	&	0.80	&	0.048	&	0.21	&	CeFe$_{10}$Pd$_{2}$	&	1.03	&	23.76	&	0.72	&	0.040	&	0.18	\\
CeFe$_{11}$Ag	&	1.05	&	25.30	&	0.68	&	0.020	&	-0.08	&	CeFe$_{10}$Ag$_{2}$	&	0.76	&	22.32	&	0.65	&	0.011	&	-0.09	\\
CeFe$_{11}$Cd	&	0.89	&	24.37	&	0.65	&	0.003	&	-0.25	&	CeFe$_{10}$Cd$_{2}$	&	0.83	&	20.81	&	0.73	&	0.001	&	-0.22	\\
\hline																							
\multicolumn{12}{c}{5$d$ elements}		\\																				
\hline																							
        &              MAE      &       $m$   &       $\kappa$        &       $m_l$(X)   &       $m_s$(X)	&	        &              MAE      &       $m$ &       $\kappa$        &       $m_l$(X)   &       $m_s$(X)	\\																				
																							
\hline																							
CeFe$_{11}$Lu	&	0.02	&	22.08	&	0.11	&	0.013	&	-0.41	&	CeFe$_{10}$Lu$_{2}$	&	-0.51	&	19.18	&	0.00	&	0.020	&	-0.41	\\
CeFe$_{11}$Hf	&	0.36	&	21.85	&	0.46	&	0.029	&	-0.74	&	CeFe$_{10}$Hf$_{2}$	&	0.91	&	18.58	&	0.86	&	0.014	&	-0.52	\\
CeFe$_{11}$Ta	&	1.66	&	21.88	&	0.99	&	0.041	&	-1.02	&	CeFe$_{10}$Ta$_{2}$	&	2.04	&	17.97	&	1.33	&	0.020	&	-0.61	\\
CeFe$_{11}$W	&	1.47	&	22.09	&	0.92	&	0.015	&	-0.95	&	CeFe$_{10}$W$_{2}$	&	2.44	&	18.26	&	1.43	&	0.010	&	-0.59	\\
CeFe$_{11}$Re	&	2.01	&	22.00	&	1.08	&	-0.053	&	-0.71	&	CeFe$_{10}$Re$_{2}$	&	0.52	&	18.56	&	0.65	&	-0.027	&	-0.46	\\
CeFe$_{11}$Os	&	1.08	&	23.56	&	0.74	&	-0.160	&	-0.08	&	CeFe$_{10}$Os$_{2}$	&	-0.06	&	20.67	&	0.00	&	-0.095	&	-0.16	\\
CeFe$_{11}$Ir	&	-0.06	&	25.26	&	0.00	&	-0.036	&	0.29	&	CeFe$_{10}$Ir$_{2}$	&	-1.14	&	24.02	&	0.00	&	-0.017	&	0.35	\\
CeFe$_{11}$Pt	&	1.80	&	25.89	&	0.87	&	0.079	&	0.25	&	CeFe$_{10}$Pt$_{2}$	&	0.43	&	24.24	&	0.45	&	0.081	&	0.24	\\
CeFe$_{11}$Au	&	0.99	&	25.67	&	0.65	&	0.069	&	0.01	&	CeFe$_{10}$Au$_{2}$	&	0.02	&	22.97	&	0.10	&	0.047	&	-0.02	\\
CeFe$_{11}$Hg	&	0.55	&	24.76	&	0.50	&	0.027	&	-0.17	&	CeFe$_{10}$Hg$_{2}$	&	0.97	&	21.36	&	0.77	&	0.015	&	-0.16	\\

\end{tabular}
\end{ruledtabular}
\end{table*}

\section*{ACKNOWLEDGMENTS}
We gratefully acknowledge financial support from the National Science Center Poland under decisions DEC-2019/35/O/ST5/02980 (PRELUDIUM-BIS 1) and DEC-2018/30/E/ST3/00267 (SONATA-BIS 8).
We thank Paweł Leśniak and Daniel Depcik for compiling the scientific software and administering the computational cluster at the Institute of Molecular Physics, Polish Academy of Sciences.

\section*{APPENDIX}
The lattice parameters and atomic positions of the considered systems "1-11-1" and "1-10-2" and the compound CeFe$_{12}$ are presented in Tables~\ref{tab:Pmmn_I4mmm} and ~\ref{tab:Combinations}. The presented parameters relate directly to the crystallographic structures ilustrated in Figs.~\ref{fig:CeFe10W2_P4mmm} and \ref{fig:fig1}.  In the present work the structures shown in Table~\ref{tab:Pmmn_I4mmm} are used. CeFe$_{11}$TiH structure is shown for the comparision as an experimental result~\citep{isnard_hydrogen_1998}.

\begin{table*}
\caption{\label{tab:Pmmn_I4mmm} Lattice parameters (in \AA{}) and atomic positions for CeFe$_{10}$X$_{2}$ (s.g.~$P$4/$mmm$, No.~123), CeFe$_{11}$Ti (s.g.~{\it Pmmn}, No.~59), CeFe$_{12}$ (s.g.~$I$4/$mmm$, No.~139), and CeFe$_{11}$TiH (s.g.~$I$4/$mmm$, No.~139). See Fig.~\ref{fig:fig1} for illustrations of the structures. The atomic positions presented for CeFe$_{11}$TiH were obtained experimentally by Isnard $et al.$~\citep{isnard_hydrogen_1998}.
}
\begin{tabular}{llll|llll|llll|llll }
 \hline
 \hline
 \multicolumn{4}{c}{CeFe$_{10}$X$_{2}$ s.g. $P$4/$mmm$}  & \multicolumn{4}{c}{CeFe$_{11}$Ti s.g. $Pmmn$} & \multicolumn{4}{c}{CeFe$_{12}$ s.g. $I$4/$mmm$} & \multicolumn{4}{c}{CeFe$_{11}$TiH s.g. $I$4/$mmm$}\\
 \hline
$a,b,c$& 8.539 &     8.539   &    4.78   &   $a,b,c$      &  8.539 &   4.78 &       8.539     &   $a,b,c$     & 8.539 &     8.539   &    4.78 & $a,b,c$& 8.566 &     8.566 &    4.80   \\
Site &	  ~$x$ 	&	   $y$ 	&	       $z$      & Site 	&	  $x$ 	&	   $y$ 	&	       $z$      & Site 	&	  ~$x$ 	&	   $y$ 	&	       $z$ & Site &	  ~$x$ 	&	   $y$ 	&	       $z$      \\
 \hline
Ce 	& ~0.0 & ~0.0 & ~0.0 & Ce & 0.25 & 0.25 & 0.25			&	  Ce &  ~0.0 & 0.0 & 0.0   & Ce (2$a$) & ~0.0 & ~0.0 & ~0.0 \\
Ce 	& -0.5 & -0.5 & -0.5  & Fe & 0.5    & 0.5  & 0.5			&	  Fe &  ~0.3582 & 0.0 & 0.0  & Fe (8$i$) & ~0.3534~(6) & ~0.0 & ~0.0 \\
Fe 	& ~0.251 & ~0.251 & ~0.254  & Fe & 0.5    & 0.5   & 0.0		&Fe 	& -0.2275 & 0.0  & 0.5   	& Ti (8$i$) & ~0.3534~(6) & ~0.0 & ~0.0 \\
Ti & ~0.349 & ~0.0 & ~0.0 & Fe & 0.89   & 0.25 & 0.25			&	  Fe &  -0.25 & 0.25  & 0.25 & Fe (8$j$) & ~0.2753~(4) & ~0.5 & ~0.0 \\
Fe 	& -0.141 & -0.5 & -0.5    & Fe & 0.25   & 0.25 & 0.89			&	     &        &      &       & Fe (8$f$) & ~0.25 & ~0.25 & ~0.25 \\
Fe 	& ~0.277 & -0.5 & ~0.0    & Fe & 0.75 & 0.25  & 0.515			&	     &        &      &       & H (2$b$)  & ~0.0 & ~0.0 & ~0.5 \\ 	
Fe 	& -0.219 & ~0.0 & -0.5    & Fe & 0.75 & 0.25 & 0.985			&	     &        &      &       &\\	
	& & &   & Fe & 0.515 & 0.25 & 0.75			&	     &        &       & \\	
	& & &   & Ti & 0.25 & 0.25 & 0.61			&	     &        &       & \\	

\hline 
\hline
\end{tabular}
\end{table*}
\end{sloppypar}
\begin{table*}
\caption{\label{tab:Combinations} 
Lattice parameters (in \AA{}) and atomic positions for several considered crystal structures of CeFe$_{11}$Ti with space groups $Amm$2 (No. 38), $Ama$2 (No. 40), $Imm$2 (No. 44), and $Pmmm$ (No. 47). See Fig.~\ref{fig:fig1} for illustrations of the structures.
}
\begin{tabular}{llll|llll|llll|llll}
 \hline
 \hline
 \multicolumn{4}{c}{$Amm$2} & \multicolumn{4}{c}{$Ama$2} & \multicolumn{4}{c}{$Imm$2} & \multicolumn{4}{c}{$Pmmm$} \\
 \hline

$a,b,c$		&	4.78	&	12.075	&	12.075	&	   $a,b,c$   	&	4.78	&	12.075	&	12.075	&	 $a,b,c$	&	4.78	&	8.539	&	8.539	&	      $a,b,c$	&	8.539	&	8.539	&	4.78 \\
	 site 	&	  $x$ 	&	   $y$ 	&	       $z$     	&	 	 site 	&	  $x$ 	&	   $y$ 	&	       $z$ 	&		 site 	&	  $x$ 	&	   $y$ 	&	       $z$      	&	 	 site 	&	  $x$ 	&	   $y$ 	&	       $z$  \\
\hline																															
	        Ce     	&	0	&	0	&	0	&	        Ce     	&	0.25	&	0.75	&	0.25	&	         Ce    	&	0	&	0	&	0	&	          Ce           	&	0.5	&	0.5	&	       0.5     \\
	        Ce     	&	0.5	&	0.5	&	0	&	        Fe     	&	0.5	&	0.25	&	0	&	        Fe     	&	0.75	&	0.75	&	0.0025	&	         Ce    	&	0	&	0	&	       0       \\
	        Fe     	&	0.25	&	0.75	&	0	&	        Fe     	&	0.5	&	0.5	&	0.25	&	         Fe    	&	0	&	0.64	&	0	&	         Fe            	&	0.75	&	0.75	&	       0.75    \\
	        Fe     	&	0.25	&	0.5	&	0.25	&	        Fe     	&	0.5	&	0	&	0.25	&	         Fe    	&	0	&	0	&	0.36	&	         Fe    	&	0.5	&	0.14	&	       0.5     \\
	        Fe     	&	0.25	&	0	&	0.25	&	        Fe     	&	0.25	&	0.93	&	0.07	&	         Fe    	&	0.5	&	0	&	0.235	&	         Fe    	&	0.36	&	0	&	       0       \\
	        Fe     	&	0	&	0.32	&	0.32	&	        Fe     	&	0.25	&	0.57	&	0.07	&	        Fe     	&	0	&	0.5	&	0.265	&	         Fe    	&	0	&	0.36	&	       0       \\
	        Fe     	&	0.5	&	0.32	&	0.18	&	        Fe     	&	0.25	&	0.57	&	0.43	&	        Fe     	&	0	&	0.265	&	0.5	&	         Fe    	&	0.765	&	0	&	       0.5     \\
	        Fe     	&	0.5	&	0.18	&	0.32	&	        Fe     	&	0.25	&	0.1325	&	0.1325	&	        Ti     	&	0	&	0	&	0.64	&	         Fe    	&	0	&	0.765	&	       0.5     \\
	        Fe     	&	0	&	0.1175	&	0.3825	&	        Fe     	&	0.25	&	0.3675	&	0.3675	&	               	&	               	&	               	&	                                       	&	         Fe    	&	0.265	&	0.5	&	       0       \\
	        Fe     	&	0	&	0.3825	&	0.1175	&	        Fe     	&	0.25	&	0.3675	&	0.1325	&	               	&	               	&	               	&	                                       	&	        Fe     	&	0.5	&	0.265	&	       0       \\
	        Fe     	&	0.5	&	0.6175	&	0.3825	&	        Fe     	&	0.25	&	0.1325	&	0.3675	&	               	&	               	&	               	&	                                       	&	        Ti     	&	0.86	&	0.5	&	       0.5     \\
	        Fe     	&	0.5	&	0.8825	&	0.1175	&	        Ti     	&	0.25	&	0.93	&	0.43	&	               	&	               	&	               	&	               	&	               	&	               	&	               	&	               \\
	        Ti     	&	0	&	0.82	&	0.18	&	               	&	               	&	                       	&	                       	&	               	&	               	&	               	&	               	&	               	&	               	&	               	&	               \\
				
\hline 
\hline
\end{tabular}
\end{table*}

\break
\bibliography{CeFe11Ti}

\begin{thebibliography}{49}%
\makeatletter
\providecommand \@ifxundefined [1]{%
 \@ifx{#1\undefined}
}%
\providecommand \@ifnum [1]{%
 \ifnum #1\expandafter \@firstoftwo
 \else \expandafter \@secondoftwo
 \fi
}%
\providecommand \@ifx [1]{%
 \ifx #1\expandafter \@firstoftwo
 \else \expandafter \@secondoftwo
 \fi
}%
\providecommand \natexlab [1]{#1}%
\providecommand \enquote  [1]{``#1''}%
\providecommand \bibnamefont  [1]{#1}%
\providecommand \bibfnamefont [1]{#1}%
\providecommand \citenamefont [1]{#1}%
\providecommand \href@noop [0]{\@secondoftwo}%
\providecommand \href [0]{\begingroup \@sanitize@url \@href}%
\providecommand \@href[1]{\@@startlink{#1}\@@href}%
\providecommand \@@href[1]{\endgroup#1\@@endlink}%
\providecommand \@sanitize@url [0]{\catcode `\\12\catcode `\$12\catcode
  `\&12\catcode `\#12\catcode `\^12\catcode `\_12\catcode `\%12\relax}%
\providecommand \@@startlink[1]{}%
\providecommand \@@endlink[0]{}%
\providecommand \url  [0]{\begingroup\@sanitize@url \@url }%
\providecommand \@url [1]{\endgroup\@href {#1}{\urlprefix }}%
\providecommand \urlprefix  [0]{URL }%
\providecommand \Eprint [0]{\href }%
\providecommand \doibase [0]{https://doi.org/}%
\providecommand \selectlanguage [0]{\@gobble}%
\providecommand \bibinfo  [0]{\@secondoftwo}%
\providecommand \bibfield  [0]{\@secondoftwo}%
\providecommand \translation [1]{[#1]}%
\providecommand \BibitemOpen [0]{}%
\providecommand \bibitemStop [0]{}%
\providecommand \bibitemNoStop [0]{.\EOS\space}%
\providecommand \EOS [0]{\spacefactor3000\relax}%
\providecommand \BibitemShut  [1]{\csname bibitem#1\endcsname}%
\let\auto@bib@innerbib\@empty
\bibitem [{\citenamefont {Toga}\ \emph {et~al.}(2018)\citenamefont {Toga},
  \citenamefont {Nishino}, \citenamefont {Miyashita}, \citenamefont {Miyake},\
  and\ \citenamefont {Sakuma}}]{toga_anisotropy_2018}%
  \BibitemOpen
  \bibfield  {author} {\bibinfo {author} {\bibfnamefont {Y.}~\bibnamefont
  {Toga}}, \bibinfo {author} {\bibfnamefont {M.}~\bibnamefont {Nishino}},
  \bibinfo {author} {\bibfnamefont {S.}~\bibnamefont {Miyashita}}, \bibinfo
  {author} {\bibfnamefont {T.}~\bibnamefont {Miyake}},\ and\ \bibinfo {author}
  {\bibfnamefont {A.}~\bibnamefont {Sakuma}},\ }\bibfield  {title} {\bibinfo
  {title} {Anisotropy of exchange stiffness based on atomic-scale magnetic
  properties in the rare-earth permanent magnet
  {{Nd}}{\textsubscript{2}}{{Fe}}{\textsubscript{14}}{{B}}},\ }\href
  {https://doi.org/10.1103/PhysRevB.98.054418} {\bibfield  {journal} {\bibinfo
  {journal} {Phys. Rev. B}\ }\textbf {\bibinfo {volume} {98}},\ \bibinfo
  {pages} {054418} (\bibinfo {year} {2018})}\BibitemShut {NoStop}%
\bibitem [{\citenamefont {Das}\ \emph {et~al.}(2019)\citenamefont {Das},
  \citenamefont {Choudhary}, \citenamefont {Skomski}, \citenamefont
  {Balasubramanian}, \citenamefont {Pathak}, \citenamefont {Paudyal},\ and\
  \citenamefont {Sellmyer}}]{das_anisotropy_2019}%
  \BibitemOpen
  \bibfield  {author} {\bibinfo {author} {\bibfnamefont {B.}~\bibnamefont
  {Das}}, \bibinfo {author} {\bibfnamefont {R.}~\bibnamefont {Choudhary}},
  \bibinfo {author} {\bibfnamefont {R.}~\bibnamefont {Skomski}}, \bibinfo
  {author} {\bibfnamefont {B.}~\bibnamefont {Balasubramanian}}, \bibinfo
  {author} {\bibfnamefont {A.~K.}\ \bibnamefont {Pathak}}, \bibinfo {author}
  {\bibfnamefont {D.}~\bibnamefont {Paudyal}},\ and\ \bibinfo {author}
  {\bibfnamefont {D.~J.}\ \bibnamefont {Sellmyer}},\ }\bibfield  {title}
  {\bibinfo {title} {Anisotropy and orbital moment in {{Sm-Co}} permanent
  magnets},\ }\href {https://doi.org/10.1103/PhysRevB.100.024419} {\bibfield
  {journal} {\bibinfo  {journal} {Phys. Rev. B}\ }\textbf {\bibinfo {volume}
  {100}},\ \bibinfo {pages} {024419} (\bibinfo {year} {2019})}\BibitemShut
  {NoStop}%
\bibitem [{\citenamefont {Bourzac}(2011)}]{bourzac_rare-earth_2011}%
  \BibitemOpen
  \bibfield  {author} {\bibinfo {author} {\bibfnamefont {K.}~\bibnamefont
  {Bourzac}},\ }\bibfield  {title} {\bibinfo {title} {The rare-earth crisis},\
  }\href@noop {} {\bibfield  {journal} {\bibinfo  {journal} {Techn. Rev.}\
  }\textbf {\bibinfo {volume} {114}},\ \bibinfo {pages} {58} (\bibinfo {year}
  {2011})}\BibitemShut {NoStop}%
\bibitem [{\citenamefont {Niarchos}\ \emph {et~al.}(2015)\citenamefont
  {Niarchos}, \citenamefont {Giannopoulos}, \citenamefont {Gjoka},
  \citenamefont {Sarafidis}, \citenamefont {Psycharis}, \citenamefont {Rusz},
  \citenamefont {Edstr{\"o}m}, \citenamefont {Eriksson}, \citenamefont {Toson},
  \citenamefont {Fidler}, \citenamefont {Anagnostopoulou}, \citenamefont
  {Sanyal}, \citenamefont {Ott}, \citenamefont {Lacroix}, \citenamefont {Viau},
  \citenamefont {Bran}, \citenamefont {Vazquez}, \citenamefont {Reichel},
  \citenamefont {Schultz},\ and\ \citenamefont
  {F{\"a}hler}}]{niarchos_toward_2015}%
  \BibitemOpen
  \bibfield  {author} {\bibinfo {author} {\bibfnamefont {D.}~\bibnamefont
  {Niarchos}}, \bibinfo {author} {\bibfnamefont {G.}~\bibnamefont
  {Giannopoulos}}, \bibinfo {author} {\bibfnamefont {M.}~\bibnamefont {Gjoka}},
  \bibinfo {author} {\bibfnamefont {C.}~\bibnamefont {Sarafidis}}, \bibinfo
  {author} {\bibfnamefont {V.}~\bibnamefont {Psycharis}}, \bibinfo {author}
  {\bibfnamefont {J.}~\bibnamefont {Rusz}}, \bibinfo {author} {\bibfnamefont
  {A.}~\bibnamefont {Edstr{\"o}m}}, \bibinfo {author} {\bibfnamefont
  {O.}~\bibnamefont {Eriksson}}, \bibinfo {author} {\bibfnamefont
  {P.}~\bibnamefont {Toson}}, \bibinfo {author} {\bibfnamefont
  {J.}~\bibnamefont {Fidler}}, \bibinfo {author} {\bibfnamefont
  {E.}~\bibnamefont {Anagnostopoulou}}, \bibinfo {author} {\bibfnamefont
  {U.}~\bibnamefont {Sanyal}}, \bibinfo {author} {\bibfnamefont
  {F.}~\bibnamefont {Ott}}, \bibinfo {author} {\bibfnamefont {L.-M.}\
  \bibnamefont {Lacroix}}, \bibinfo {author} {\bibfnamefont {G.}~\bibnamefont
  {Viau}}, \bibinfo {author} {\bibfnamefont {C.}~\bibnamefont {Bran}}, \bibinfo
  {author} {\bibfnamefont {M.}~\bibnamefont {Vazquez}}, \bibinfo {author}
  {\bibfnamefont {L.}~\bibnamefont {Reichel}}, \bibinfo {author} {\bibfnamefont
  {L.}~\bibnamefont {Schultz}},\ and\ \bibinfo {author} {\bibfnamefont
  {S.}~\bibnamefont {F{\"a}hler}},\ }\bibfield  {title} {\bibinfo {title}
  {Toward {{Rare-Earth-Free Permanent Magnets}}: {{A Combinatorial Approach
  Exploiting}} the {{Possibilities}} of {{Modeling}}, {{Shape Anisotropy}} in
  {{Elongated Nanoparticles}}, and {{Combinatorial Thin-Film Approach}}},\
  }\href {https://doi.org/10.1007/s11837-015-1431-7} {\bibfield  {journal}
  {\bibinfo  {journal} {JOM J. Miner. Met. Mater. Soc.}\ }\textbf {\bibinfo
  {volume} {67}},\ \bibinfo {pages} {1318} (\bibinfo {year}
  {2015})}\BibitemShut {NoStop}%
\bibitem [{\citenamefont {Hirosawa}(2015)}]{hirosawa_current_2015}%
  \BibitemOpen
  \bibfield  {author} {\bibinfo {author} {\bibfnamefont {S.}~\bibnamefont
  {Hirosawa}},\ }\bibfield  {title} {\bibinfo {title} {Current {{Status}} of
  {{Research}} and {{Development}} toward {{Permanent Magnets Free}} from
  {{Critical Elements}}},\ }\href {https://doi.org/10.3379/msjmag.1504R004}
  {\bibfield  {journal} {\bibinfo  {journal} {J. Magn. Soc. Jpn.}\ }\textbf
  {\bibinfo {volume} {39}},\ \bibinfo {pages} {85} (\bibinfo {year}
  {2015})}\BibitemShut {NoStop}%
\bibitem [{\citenamefont {Li}\ \emph {et~al.}(2015)\citenamefont {Li},
  \citenamefont {Pan}, \citenamefont {Li},\ and\ \citenamefont
  {Zhang}}]{li_recent_2015}%
  \BibitemOpen
  \bibfield  {author} {\bibinfo {author} {\bibfnamefont {D.}~\bibnamefont
  {Li}}, \bibinfo {author} {\bibfnamefont {D.}~\bibnamefont {Pan}}, \bibinfo
  {author} {\bibfnamefont {S.}~\bibnamefont {Li}},\ and\ \bibinfo {author}
  {\bibfnamefont {Z.}~\bibnamefont {Zhang}},\ }\bibfield  {title} {\bibinfo
  {title} {Recent developments of rare-earth-free hard-magnetic materials},\
  }\href {https://doi.org/10.1007/s11433-015-5760-x} {\bibfield  {journal}
  {\bibinfo  {journal} {Sci. China Phys. Mech. Astron.}\ }\textbf {\bibinfo
  {volume} {59}},\ \bibinfo {pages} {617501} (\bibinfo {year}
  {2015})}\BibitemShut {NoStop}%
\bibitem [{\citenamefont {Hirosawa}\ \emph {et~al.}(2017)\citenamefont
  {Hirosawa}, \citenamefont {Nishino},\ and\ \citenamefont
  {Miyashita}}]{hirosawa_perspectives_2017}%
  \BibitemOpen
  \bibfield  {author} {\bibinfo {author} {\bibfnamefont {S.}~\bibnamefont
  {Hirosawa}}, \bibinfo {author} {\bibfnamefont {M.}~\bibnamefont {Nishino}},\
  and\ \bibinfo {author} {\bibfnamefont {S.}~\bibnamefont {Miyashita}},\
  }\bibfield  {title} {\bibinfo {title} {Perspectives for high-performance
  permanent magnets: {{Applications}}, coercivity, and new materials},\ }\href
  {https://doi.org/10.1088/2043-6254/aa597c} {\bibfield  {journal} {\bibinfo
  {journal} {Adv. Nat. Sci. Nanosci. Nanotechnol.}\ }\textbf {\bibinfo {volume}
  {8}},\ \bibinfo {pages} {013002} (\bibinfo {year} {2017})}\BibitemShut
  {NoStop}%
\bibitem [{\citenamefont {Skomski}\ and\ \citenamefont
  {Coey}(2016)}]{skomski_magnetic_2016}%
  \BibitemOpen
  \bibfield  {author} {\bibinfo {author} {\bibfnamefont {R.}~\bibnamefont
  {Skomski}}\ and\ \bibinfo {author} {\bibfnamefont {J.}~\bibnamefont {Coey}},\
  }\bibfield  {title} {\bibinfo {title} {Magnetic anisotropy \textemdash{}
  {{How}} much is enough for a permanent magnet?},\ }\href
  {https://doi.org/10.1016/j.scriptamat.2015.09.021} {\bibfield  {journal}
  {\bibinfo  {journal} {Scr. Mater.}\ }\textbf {\bibinfo {volume} {112}},\
  \bibinfo {pages} {3} (\bibinfo {year} {2016})}\BibitemShut {NoStop}%
\bibitem [{\citenamefont {Ener}\ \emph {et~al.}(2021)\citenamefont {Ener},
  \citenamefont {Skokov}, \citenamefont {Palanisamy}, \citenamefont
  {Devillers}, \citenamefont {Fischbacher}, \citenamefont {Eslava},
  \citenamefont {Maccari}, \citenamefont {Sch{\"a}fer}, \citenamefont {Diop},
  \citenamefont {Radulov}, \citenamefont {Gault}, \citenamefont {Hrkac},
  \citenamefont {Dempsey}, \citenamefont {Schrefl}, \citenamefont {Raabe},\
  and\ \citenamefont {Gutfleisch}}]{ener_twins_2021}%
  \BibitemOpen
  \bibfield  {author} {\bibinfo {author} {\bibfnamefont {S.}~\bibnamefont
  {Ener}}, \bibinfo {author} {\bibfnamefont {K.~P.}\ \bibnamefont {Skokov}},
  \bibinfo {author} {\bibfnamefont {D.}~\bibnamefont {Palanisamy}}, \bibinfo
  {author} {\bibfnamefont {T.}~\bibnamefont {Devillers}}, \bibinfo {author}
  {\bibfnamefont {J.}~\bibnamefont {Fischbacher}}, \bibinfo {author}
  {\bibfnamefont {G.~G.}\ \bibnamefont {Eslava}}, \bibinfo {author}
  {\bibfnamefont {F.}~\bibnamefont {Maccari}}, \bibinfo {author} {\bibfnamefont
  {L.}~\bibnamefont {Sch{\"a}fer}}, \bibinfo {author} {\bibfnamefont
  {L.~V.~B.}\ \bibnamefont {Diop}}, \bibinfo {author} {\bibfnamefont
  {I.}~\bibnamefont {Radulov}}, \bibinfo {author} {\bibfnamefont
  {B.}~\bibnamefont {Gault}}, \bibinfo {author} {\bibfnamefont
  {G.}~\bibnamefont {Hrkac}}, \bibinfo {author} {\bibfnamefont {N.~M.}\
  \bibnamefont {Dempsey}}, \bibinfo {author} {\bibfnamefont {T.}~\bibnamefont
  {Schrefl}}, \bibinfo {author} {\bibfnamefont {D.}~\bibnamefont {Raabe}},\
  and\ \bibinfo {author} {\bibfnamefont {O.}~\bibnamefont {Gutfleisch}},\
  }\bibfield  {title} {\bibinfo {title} {Twins \textendash{} {{A}} weak link in
  the magnetic hardening of {{ThMn}}{\textsubscript{12}}-type permanent
  magnets},\ }\href {https://doi.org/10.1016/j.actamat.2021.116968} {\bibfield
  {journal} {\bibinfo  {journal} {Acta Mater.}\ ,\ \bibinfo {pages} {116968}}
  (\bibinfo {year} {2021})}\BibitemShut {NoStop}%
\bibitem [{\citenamefont {Werwi{\'n}ski}\ and\ \citenamefont
  {Marciniak}(2017)}]{werwinski_ab_2017-1}%
  \BibitemOpen
  \bibfield  {author} {\bibinfo {author} {\bibfnamefont {M.}~\bibnamefont
  {Werwi{\'n}ski}}\ and\ \bibinfo {author} {\bibfnamefont {W.}~\bibnamefont
  {Marciniak}},\ }\bibfield  {title} {\bibinfo {title} {Ab initio study of
  magnetocrystalline anisotropy, magnetostriction, and {{Fermi}} surface of
  {{L1}}{\textsubscript{0}} {{FeNi}} (tetrataenite)},\ }\href
  {https://doi.org/10.1088/1361-6463/aa958a} {\bibfield  {journal} {\bibinfo
  {journal} {J. Phys. D: Appl. Phys.}\ }\textbf {\bibinfo {volume} {50}},\
  \bibinfo {pages} {495008} (\bibinfo {year} {2017})}\BibitemShut {NoStop}%
\bibitem [{\citenamefont {Gutfleisch}\ \emph {et~al.}(2011)\citenamefont
  {Gutfleisch}, \citenamefont {Willard}, \citenamefont {Br{\"u}ck},
  \citenamefont {Chen}, \citenamefont {Sankar},\ and\ \citenamefont
  {Liu}}]{gutfleisch_magnetic_2011}%
  \BibitemOpen
  \bibfield  {author} {\bibinfo {author} {\bibfnamefont {O.}~\bibnamefont
  {Gutfleisch}}, \bibinfo {author} {\bibfnamefont {M.~A.}\ \bibnamefont
  {Willard}}, \bibinfo {author} {\bibfnamefont {E.}~\bibnamefont {Br{\"u}ck}},
  \bibinfo {author} {\bibfnamefont {C.~H.}\ \bibnamefont {Chen}}, \bibinfo
  {author} {\bibfnamefont {S.~G.}\ \bibnamefont {Sankar}},\ and\ \bibinfo
  {author} {\bibfnamefont {J.~P.}\ \bibnamefont {Liu}},\ }\bibfield  {title}
  {\bibinfo {title} {Magnetic {{Materials}} and {{Devices}} for the 21st
  {{Century}}: {{Stronger}}, {{Lighter}}, and {{More Energy Efficient}}},\
  }\href {https://doi.org/10.1002/adma.201002180} {\bibfield  {journal}
  {\bibinfo  {journal} {Adv. Mater.}\ }\textbf {\bibinfo {volume} {23}},\
  \bibinfo {pages} {821} (\bibinfo {year} {2011})}\BibitemShut {NoStop}%
\bibitem [{\citenamefont {Delange}\ \emph {et~al.}(2017)\citenamefont
  {Delange}, \citenamefont {Biermann}, \citenamefont {Miyake},\ and\
  \citenamefont {Pourovskii}}]{delange_crystal-field_2017}%
  \BibitemOpen
  \bibfield  {author} {\bibinfo {author} {\bibfnamefont {P.}~\bibnamefont
  {Delange}}, \bibinfo {author} {\bibfnamefont {S.}~\bibnamefont {Biermann}},
  \bibinfo {author} {\bibfnamefont {T.}~\bibnamefont {Miyake}},\ and\ \bibinfo
  {author} {\bibfnamefont {L.}~\bibnamefont {Pourovskii}},\ }\bibfield  {title}
  {\bibinfo {title} {Crystal-field splittings in rare-earth-based hard magnets:
  {{An}} ab initio approach},\ }\href
  {https://doi.org/10.1103/PhysRevB.96.155132} {\bibfield  {journal} {\bibinfo
  {journal} {Phys. Rev. B}\ }\textbf {\bibinfo {volume} {96}},\ \bibinfo
  {pages} {155132} (\bibinfo {year} {2017})}\BibitemShut {NoStop}%
\bibitem [{\citenamefont {S{\"o}zen}\ \emph {et~al.}(2019)\citenamefont
  {S{\"o}zen}, \citenamefont {Ener}, \citenamefont {Maccari}, \citenamefont
  {Skokov}, \citenamefont {Gutfleisch}, \citenamefont {K{\"o}rmann},
  \citenamefont {Neugebauer},\ and\ \citenamefont {Hickel}}]{sozen_ab_2019}%
  \BibitemOpen
  \bibfield  {author} {\bibinfo {author} {\bibfnamefont {H.~{\.I}.}\
  \bibnamefont {S{\"o}zen}}, \bibinfo {author} {\bibfnamefont {S.}~\bibnamefont
  {Ener}}, \bibinfo {author} {\bibfnamefont {F.}~\bibnamefont {Maccari}},
  \bibinfo {author} {\bibfnamefont {K.~P.}\ \bibnamefont {Skokov}}, \bibinfo
  {author} {\bibfnamefont {O.}~\bibnamefont {Gutfleisch}}, \bibinfo {author}
  {\bibfnamefont {F.}~\bibnamefont {K{\"o}rmann}}, \bibinfo {author}
  {\bibfnamefont {J.}~\bibnamefont {Neugebauer}},\ and\ \bibinfo {author}
  {\bibfnamefont {T.}~\bibnamefont {Hickel}},\ }\bibfield  {title} {\bibinfo
  {title} {Ab initio phase stabilities of {{Ce-based}} hard magnetic materials
  and comparison with experimental phase diagrams},\ }\href
  {https://doi.org/10.1103/PhysRevMaterials.3.084407} {\bibfield  {journal}
  {\bibinfo  {journal} {Phys. Rev. Mater.}\ }\textbf {\bibinfo {volume} {3}},\
  \bibinfo {pages} {084407} (\bibinfo {year} {2019})}\BibitemShut {NoStop}%
\bibitem [{\citenamefont {Zhou}\ and\ \citenamefont
  {Pinkerton}(2014)}]{zhou_magnetic_2014}%
  \BibitemOpen
  \bibfield  {author} {\bibinfo {author} {\bibfnamefont {C.}~\bibnamefont
  {Zhou}}\ and\ \bibinfo {author} {\bibfnamefont {F.~E.}\ \bibnamefont
  {Pinkerton}},\ }\bibfield  {title} {\bibinfo {title} {Magnetic hardening of
  {{CeFe}}{\textsubscript{12-x}}{{Mo}}{\textsubscript{x}} and the effect of
  nitrogenation},\ }\href {https://doi.org/10.1016/j.jallcom.2013.08.175}
  {\bibfield  {journal} {\bibinfo  {journal} {J. Alloys Compd.}\ }\textbf
  {\bibinfo {volume} {583}},\ \bibinfo {pages} {345} (\bibinfo {year}
  {2014})}\BibitemShut {NoStop}%
\bibitem [{\citenamefont {Hadjipanayis}\ \emph {et~al.}(2020)\citenamefont
  {Hadjipanayis}, \citenamefont {Gabay}, \citenamefont {Sch{\"o}nh{\"o}bel},
  \citenamefont {{Mart{\'i}n-Cid}}, \citenamefont {Barandiaran},\ and\
  \citenamefont {Niarchos}}]{hadjipanayis_thmn12-type_2020}%
  \BibitemOpen
  \bibfield  {author} {\bibinfo {author} {\bibfnamefont {G.}~\bibnamefont
  {Hadjipanayis}}, \bibinfo {author} {\bibfnamefont {A.}~\bibnamefont {Gabay}},
  \bibinfo {author} {\bibfnamefont {A.}~\bibnamefont {Sch{\"o}nh{\"o}bel}},
  \bibinfo {author} {\bibfnamefont {A.}~\bibnamefont {{Mart{\'i}n-Cid}}},
  \bibinfo {author} {\bibfnamefont {J.}~\bibnamefont {Barandiaran}},\ and\
  \bibinfo {author} {\bibfnamefont {D.}~\bibnamefont {Niarchos}},\ }\bibfield
  {title} {\bibinfo {title} {{{ThMn}}{\textsubscript{12}}-{{Type Alloys}} for
  {{Permanent Magnets}}},\ }\href {https://doi.org/10.1016/j.eng.2018.12.011}
  {\bibfield  {journal} {\bibinfo  {journal} {Engineering}\ }\textbf {\bibinfo
  {volume} {6}},\ \bibinfo {pages} {141} (\bibinfo {year} {2020})}\BibitemShut
  {NoStop}%
\bibitem [{\citenamefont {Pan}\ \emph {et~al.}(1994)\citenamefont {Pan},
  \citenamefont {Liu},\ and\ \citenamefont {Yang}}]{pan_structural_1994}%
  \BibitemOpen
  \bibfield  {author} {\bibinfo {author} {\bibfnamefont {Q.}~\bibnamefont
  {Pan}}, \bibinfo {author} {\bibfnamefont {Z.-X.}\ \bibnamefont {Liu}},\ and\
  \bibinfo {author} {\bibfnamefont {Y.-C.}\ \bibnamefont {Yang}},\ }\bibfield
  {title} {\bibinfo {title} {Structural and magnetic properties of
  {{Ce}}({{Fe}},{{M}}){\textsubscript{12}}{{N}}{\textsubscript{x}} interstitial
  compounds, {{M}}={{Ti}}, {{V}}, {{Cr}}, and {{Mo}}},\ }\href
  {https://doi.org/10.1063/1.358184} {\bibfield  {journal} {\bibinfo  {journal}
  {J. Appl. Phys.}\ }\textbf {\bibinfo {volume} {76}},\ \bibinfo {pages} {6728}
  (\bibinfo {year} {1994})}\BibitemShut {NoStop}%
\bibitem [{\citenamefont {Akayama}\ \emph {et~al.}(1994)\citenamefont
  {Akayama}, \citenamefont {Fujii}, \citenamefont {Yamamoto},\ and\
  \citenamefont {Tatami}}]{akayama_physical_1994}%
  \BibitemOpen
  \bibfield  {author} {\bibinfo {author} {\bibfnamefont {M.}~\bibnamefont
  {Akayama}}, \bibinfo {author} {\bibfnamefont {H.}~\bibnamefont {Fujii}},
  \bibinfo {author} {\bibfnamefont {K.}~\bibnamefont {Yamamoto}},\ and\
  \bibinfo {author} {\bibfnamefont {K.}~\bibnamefont {Tatami}},\ }\bibfield
  {title} {\bibinfo {title} {Physical properties of nitrogenated
  {{RFe}}{\textsubscript{11}}{{Ti}} intermetallic compounds ({{R}}={{Ce}},
  {{Pr}} and {{Nd}}) with {{ThMn}}{\textsubscript{12}}-type structure},\ }\href
  {https://doi.org/10.1016/0304-8853(94)90662-9} {\bibfield  {journal}
  {\bibinfo  {journal} {J. Magn. Magn. Mater.}\ }\textbf {\bibinfo {volume}
  {130}},\ \bibinfo {pages} {99} (\bibinfo {year} {1994})}\BibitemShut
  {NoStop}%
\bibitem [{\citenamefont {Isnard}\ \emph {et~al.}(1998)\citenamefont {Isnard},
  \citenamefont {Miraglia}, \citenamefont {Guillot},\ and\ \citenamefont
  {Fruchart}}]{isnard_hydrogen_1998}%
  \BibitemOpen
  \bibfield  {author} {\bibinfo {author} {\bibfnamefont {O.}~\bibnamefont
  {Isnard}}, \bibinfo {author} {\bibfnamefont {S.}~\bibnamefont {Miraglia}},
  \bibinfo {author} {\bibfnamefont {M.}~\bibnamefont {Guillot}},\ and\ \bibinfo
  {author} {\bibfnamefont {D.}~\bibnamefont {Fruchart}},\ }\bibfield  {title}
  {\bibinfo {title} {Hydrogen effects on the magnetic properties of
  {{RFe}}{\textsubscript{11}}{{Ti}} compounds},\ }\href
  {https://doi.org/10.1016/S0925-8388(98)00409-5} {\bibfield  {journal}
  {\bibinfo  {journal} {J. Alloys Compd.}\ }\textbf {\bibinfo {volume}
  {275--277}},\ \bibinfo {pages} {637} (\bibinfo {year} {1998})}\BibitemShut
  {NoStop}%
\bibitem [{\citenamefont {Ke}\ and\ \citenamefont
  {Johnson}(2016)}]{ke_intrinsic_2016-1}%
  \BibitemOpen
  \bibfield  {author} {\bibinfo {author} {\bibfnamefont {L.}~\bibnamefont
  {Ke}}\ and\ \bibinfo {author} {\bibfnamefont {D.~D.}\ \bibnamefont
  {Johnson}},\ }\bibfield  {title} {\bibinfo {title} {Intrinsic magnetic
  properties in {{R}}({{Fe}}{\textsubscript{1-x}}{{Co}}{\textsubscript{x}}
  ){\textsubscript{11}}{{Ti Z}}({{R}}={{Y}} and {{Ce}}; {{Z}}={{H}}, {{C}} ,
  and {{N}})},\ }\href {https://doi.org/10.1103/PhysRevB.94.024423} {\bibfield
  {journal} {\bibinfo  {journal} {Phys. Rev. B}\ }\textbf {\bibinfo {volume}
  {94}},\ \bibinfo {pages} {024423} (\bibinfo {year} {2016})}\BibitemShut
  {NoStop}%
\bibitem [{\citenamefont {{Martinez-Casado}}\ \emph {et~al.}(2019)\citenamefont
  {{Martinez-Casado}}, \citenamefont {Dasmahapatra}, \citenamefont {Sgroi},
  \citenamefont {{Romero-Mu{\~n}iz}}, \citenamefont {Herper}, \citenamefont
  {Vekilova}, \citenamefont {Ferrari}, \citenamefont {Pullini}, \citenamefont
  {Desmarais},\ and\ \citenamefont {Maschio}}]{martinez-casado_cefe11ti_2019}%
  \BibitemOpen
  \bibfield  {author} {\bibinfo {author} {\bibfnamefont {R.}~\bibnamefont
  {{Martinez-Casado}}}, \bibinfo {author} {\bibfnamefont {A.}~\bibnamefont
  {Dasmahapatra}}, \bibinfo {author} {\bibfnamefont {M.~F.}\ \bibnamefont
  {Sgroi}}, \bibinfo {author} {\bibfnamefont {C.}~\bibnamefont
  {{Romero-Mu{\~n}iz}}}, \bibinfo {author} {\bibfnamefont {H.~C.}\ \bibnamefont
  {Herper}}, \bibinfo {author} {\bibfnamefont {O.~Y.}\ \bibnamefont
  {Vekilova}}, \bibinfo {author} {\bibfnamefont {A.~M.}\ \bibnamefont
  {Ferrari}}, \bibinfo {author} {\bibfnamefont {D.}~\bibnamefont {Pullini}},
  \bibinfo {author} {\bibfnamefont {J.}~\bibnamefont {Desmarais}},\ and\
  \bibinfo {author} {\bibfnamefont {L.}~\bibnamefont {Maschio}},\ }\bibfield
  {title} {\bibinfo {title} {The {{CeFe}}{\textsubscript{11}}{{Ti}} permanent
  magnet: A closer look at the microstructure of the compound},\ }\href
  {https://doi.org/10.1088/1361-648X/ab4096} {\bibfield  {journal} {\bibinfo
  {journal} {J. Phys.: Condens. Matter}\ }\textbf {\bibinfo {volume} {31}},\
  \bibinfo {pages} {505505} (\bibinfo {year} {2019})}\BibitemShut {NoStop}%
\bibitem [{\citenamefont {Maccari}\ \emph {et~al.}(2021)\citenamefont
  {Maccari}, \citenamefont {Ener}, \citenamefont {Koch}, \citenamefont {Dirba},
  \citenamefont {Skokov}, \citenamefont {Bruder}, \citenamefont {Sch{\"a}fer},\
  and\ \citenamefont {Gutfleisch}}]{maccari_correlating_2021}%
  \BibitemOpen
  \bibfield  {author} {\bibinfo {author} {\bibfnamefont {F.}~\bibnamefont
  {Maccari}}, \bibinfo {author} {\bibfnamefont {S.}~\bibnamefont {Ener}},
  \bibinfo {author} {\bibfnamefont {D.}~\bibnamefont {Koch}}, \bibinfo {author}
  {\bibfnamefont {I.}~\bibnamefont {Dirba}}, \bibinfo {author} {\bibfnamefont
  {K.~P.}\ \bibnamefont {Skokov}}, \bibinfo {author} {\bibfnamefont
  {E.}~\bibnamefont {Bruder}}, \bibinfo {author} {\bibfnamefont
  {L.}~\bibnamefont {Sch{\"a}fer}},\ and\ \bibinfo {author} {\bibfnamefont
  {O.}~\bibnamefont {Gutfleisch}},\ }\bibfield  {title} {\bibinfo {title}
  {Correlating changes of the unit cell parameters and microstructure with
  magnetic properties in the {{CeFe}}{\textsubscript{11}}{{Ti}} compound},\
  }\href {https://doi.org/10.1016/j.jallcom.2021.158805} {\bibfield  {journal}
  {\bibinfo  {journal} {J. Alloys Compd.}\ }\textbf {\bibinfo {volume} {867}},\
  \bibinfo {pages} {158805} (\bibinfo {year} {2021})}\BibitemShut {NoStop}%
\bibitem [{\citenamefont {Tran}\ \emph {et~al.}(2014)\citenamefont {Tran},
  \citenamefont {Karsai},\ and\ \citenamefont {Blaha}}]{tran_nonmagnetic_2014}%
  \BibitemOpen
  \bibfield  {author} {\bibinfo {author} {\bibfnamefont {F.}~\bibnamefont
  {Tran}}, \bibinfo {author} {\bibfnamefont {F.}~\bibnamefont {Karsai}},\ and\
  \bibinfo {author} {\bibfnamefont {P.}~\bibnamefont {Blaha}},\ }\bibfield
  {title} {\bibinfo {title} {Nonmagnetic and ferromagnetic fcc cerium studied
  with one-electron methods},\ }\href
  {https://doi.org/10.1103/PhysRevB.89.155106} {\bibfield  {journal} {\bibinfo
  {journal} {Phys. Rev. B}\ }\textbf {\bibinfo {volume} {89}},\ \bibinfo
  {pages} {155106} (\bibinfo {year} {2014})}\BibitemShut {NoStop}%
\bibitem [{\citenamefont {Skokowski}\ \emph {et~al.}(2020)\citenamefont
  {Skokowski}, \citenamefont {Synoradzki}, \citenamefont {Werwi{\'n}ski},
  \citenamefont {Toli{\'n}ski}, \citenamefont {Bajorek},\ and\ \citenamefont
  {Che{\l}kowska}}]{skokowski_influence_2020}%
  \BibitemOpen
  \bibfield  {author} {\bibinfo {author} {\bibfnamefont {P.}~\bibnamefont
  {Skokowski}}, \bibinfo {author} {\bibfnamefont {K.}~\bibnamefont
  {Synoradzki}}, \bibinfo {author} {\bibfnamefont {M.}~\bibnamefont
  {Werwi{\'n}ski}}, \bibinfo {author} {\bibfnamefont {T.}~\bibnamefont
  {Toli{\'n}ski}}, \bibinfo {author} {\bibfnamefont {A.}~\bibnamefont
  {Bajorek}},\ and\ \bibinfo {author} {\bibfnamefont {G.}~\bibnamefont
  {Che{\l}kowska}},\ }\bibfield  {title} {\bibinfo {title} {Influence of {{Pr}}
  substitution on the physical properties of the
  {{Ce}}{\textsubscript{1-x}}{{Pr}}{\textsubscript{x}}{{CoGe}}{\textsubscript{3}}
  system: {{Combined}} experimental and first-principles study},\ }\href
  {https://doi.org/10.1103/PhysRevB.102.245127} {\bibfield  {journal} {\bibinfo
   {journal} {Phys. Rev. B}\ }\textbf {\bibinfo {volume} {102}},\ \bibinfo
  {pages} {245127} (\bibinfo {year} {2020})}\BibitemShut {NoStop}%
\bibitem [{\citenamefont {Coey}(2012)}]{coey_permanent_2012}%
  \BibitemOpen
  \bibfield  {author} {\bibinfo {author} {\bibfnamefont {J.~M.~D.}\
  \bibnamefont {Coey}},\ }\bibfield  {title} {\bibinfo {title} {Permanent
  magnets: {{Plugging}} the gap},\ }\href
  {https://doi.org/10.1016/j.scriptamat.2012.04.036} {\bibfield  {journal}
  {\bibinfo  {journal} {Scr. Mater.}\ }\textbf {\bibinfo {volume} {67}},\
  \bibinfo {pages} {524} (\bibinfo {year} {2012})}\BibitemShut {NoStop}%
\bibitem [{\citenamefont {K{\"o}rner}\ \emph {et~al.}(2016)\citenamefont
  {K{\"o}rner}, \citenamefont {Krugel},\ and\ \citenamefont
  {Els{\"a}sser}}]{korner_theoretical_2016}%
  \BibitemOpen
  \bibfield  {author} {\bibinfo {author} {\bibfnamefont {W.}~\bibnamefont
  {K{\"o}rner}}, \bibinfo {author} {\bibfnamefont {G.}~\bibnamefont {Krugel}},\
  and\ \bibinfo {author} {\bibfnamefont {C.}~\bibnamefont {Els{\"a}sser}},\
  }\bibfield  {title} {\bibinfo {title} {Theoretical screening of intermetallic
  {{ThMn}}{\textsubscript{12}}-type phases for new hard-magnetic compounds with
  low rare earth content},\ }\href {https://doi.org/10.1038/srep24686}
  {\bibfield  {journal} {\bibinfo  {journal} {Sci. Rep.}\ }\textbf {\bibinfo
  {volume} {6}},\ \bibinfo {pages} {24686} (\bibinfo {year}
  {2016})}\BibitemShut {NoStop}%
\bibitem [{\citenamefont {Koepernik}\ and\ \citenamefont
  {Eschrig}(1999)}]{koepernik_full-potential_1999-1}%
  \BibitemOpen
  \bibfield  {author} {\bibinfo {author} {\bibfnamefont {K.}~\bibnamefont
  {Koepernik}}\ and\ \bibinfo {author} {\bibfnamefont {H.}~\bibnamefont
  {Eschrig}},\ }\bibfield  {title} {\bibinfo {title} {Full-potential
  nonorthogonal local-orbital minimum-basis band-structure scheme},\ }\href
  {https://doi.org/10.1103/PhysRevB.59.1743} {\bibfield  {journal} {\bibinfo
  {journal} {Phys. Rev. B}\ }\textbf {\bibinfo {volume} {59}},\ \bibinfo
  {pages} {1743} (\bibinfo {year} {1999})}\BibitemShut {NoStop}%
\bibitem [{\citenamefont {Perdew}\ \emph {et~al.}(1996)\citenamefont {Perdew},
  \citenamefont {Burke},\ and\ \citenamefont
  {Ernzerhof}}]{perdew_generalized_1996-1}%
  \BibitemOpen
  \bibfield  {author} {\bibinfo {author} {\bibfnamefont {J.~P.}\ \bibnamefont
  {Perdew}}, \bibinfo {author} {\bibfnamefont {K.}~\bibnamefont {Burke}},\ and\
  \bibinfo {author} {\bibfnamefont {M.}~\bibnamefont {Ernzerhof}},\ }\bibfield
  {title} {\bibinfo {title} {Generalized {{Gradient Approximation Made
  Simple}}},\ }\href {https://doi.org/10.1103/PhysRevLett.77.3865} {\bibfield
  {journal} {\bibinfo  {journal} {Phys. Rev. Lett.}\ }\textbf {\bibinfo
  {volume} {77}},\ \bibinfo {pages} {3865} (\bibinfo {year}
  {1996})}\BibitemShut {NoStop}%
\bibitem [{\citenamefont {von Barth}\ and\ \citenamefont
  {Hedin}(1972)}]{barth_local_1972}%
  \BibitemOpen
  \bibfield  {author} {\bibinfo {author} {\bibfnamefont {U.}~\bibnamefont {von
  Barth}}\ and\ \bibinfo {author} {\bibfnamefont {L.}~\bibnamefont {Hedin}},\
  }\bibfield  {title} {\bibinfo {title} {A local exchange-correlation potential
  for the spin polarized case. i},\ }\href
  {https://doi.org/10.1088/0022-3719/5/13/012} {\bibfield  {journal} {\bibinfo
  {journal} {J. Phys. C: Solid State Phys.}\ }\textbf {\bibinfo {volume} {5}},\
  \bibinfo {pages} {1629} (\bibinfo {year} {1972})}\BibitemShut {NoStop}%
\bibitem [{\citenamefont {Perdew}\ and\ \citenamefont
  {Zunger}(1981)}]{perdew_self-interaction_1981}%
  \BibitemOpen
  \bibfield  {author} {\bibinfo {author} {\bibfnamefont {J.~P.}\ \bibnamefont
  {Perdew}}\ and\ \bibinfo {author} {\bibfnamefont {A.}~\bibnamefont
  {Zunger}},\ }\bibfield  {title} {\bibinfo {title} {Self-interaction
  correction to density-functional approximations for many-electron systems},\
  }\href {https://doi.org/10.1103/PhysRevB.23.5048} {\bibfield  {journal}
  {\bibinfo  {journal} {Phys. Rev. B}\ }\textbf {\bibinfo {volume} {23}},\
  \bibinfo {pages} {5048} (\bibinfo {year} {1981})}\BibitemShut {NoStop}%
\bibitem [{\citenamefont {Perdew}\ and\ \citenamefont
  {Wang}(1992)}]{perdew_accurate_1992-1}%
  \BibitemOpen
  \bibfield  {author} {\bibinfo {author} {\bibfnamefont {J.~P.}\ \bibnamefont
  {Perdew}}\ and\ \bibinfo {author} {\bibfnamefont {Y.}~\bibnamefont {Wang}},\
  }\bibfield  {title} {\bibinfo {title} {Accurate and simple analytic
  representation of the electron-gas correlation energy},\ }\href
  {https://doi.org/10.1103/PhysRevB.45.13244} {\bibfield  {journal} {\bibinfo
  {journal} {Phys. Rev. B}\ }\textbf {\bibinfo {volume} {45}},\ \bibinfo
  {pages} {13244} (\bibinfo {year} {1992})}\BibitemShut {NoStop}%
\bibitem [{\citenamefont {Edstr{\"o}m}\ \emph {et~al.}(2015)\citenamefont
  {Edstr{\"o}m}, \citenamefont {Werwi{\'n}ski}, \citenamefont {Iu{\c s}an},
  \citenamefont {Rusz}, \citenamefont {Eriksson}, \citenamefont {Skokov},
  \citenamefont {Radulov}, \citenamefont {Ener}, \citenamefont {Kuz'min},
  \citenamefont {Hong}, \citenamefont {Fries}, \citenamefont {Karpenkov},
  \citenamefont {Gutfleisch}, \citenamefont {Toson},\ and\ \citenamefont
  {Fidler}}]{edstrom_magnetic_2015-2}%
  \BibitemOpen
  \bibfield  {author} {\bibinfo {author} {\bibfnamefont {A.}~\bibnamefont
  {Edstr{\"o}m}}, \bibinfo {author} {\bibfnamefont {M.}~\bibnamefont
  {Werwi{\'n}ski}}, \bibinfo {author} {\bibfnamefont {D.}~\bibnamefont {Iu{\c
  s}an}}, \bibinfo {author} {\bibfnamefont {J.}~\bibnamefont {Rusz}}, \bibinfo
  {author} {\bibfnamefont {O.}~\bibnamefont {Eriksson}}, \bibinfo {author}
  {\bibfnamefont {K.~P.}\ \bibnamefont {Skokov}}, \bibinfo {author}
  {\bibfnamefont {I.~A.}\ \bibnamefont {Radulov}}, \bibinfo {author}
  {\bibfnamefont {S.}~\bibnamefont {Ener}}, \bibinfo {author} {\bibfnamefont
  {M.~D.}\ \bibnamefont {Kuz'min}}, \bibinfo {author} {\bibfnamefont
  {J.}~\bibnamefont {Hong}}, \bibinfo {author} {\bibfnamefont {M.}~\bibnamefont
  {Fries}}, \bibinfo {author} {\bibfnamefont {D.~Y.}\ \bibnamefont
  {Karpenkov}}, \bibinfo {author} {\bibfnamefont {O.}~\bibnamefont
  {Gutfleisch}}, \bibinfo {author} {\bibfnamefont {P.}~\bibnamefont {Toson}},\
  and\ \bibinfo {author} {\bibfnamefont {J.}~\bibnamefont {Fidler}},\
  }\bibfield  {title} {\bibinfo {title} {Magnetic properties of
  ({{Fe}}{\textsubscript{1-x}}{{Co}}{\textsubscript{x}}){\textsubscript{2}}{{B}}
  alloys and the effect of doping by 5d elements},\ }\href
  {https://doi.org/10.1103/PhysRevB.92.174413} {\bibfield  {journal} {\bibinfo
  {journal} {Phys. Rev. B}\ }\textbf {\bibinfo {volume} {92}},\ \bibinfo
  {pages} {174413} (\bibinfo {year} {2015})}\BibitemShut {NoStop}%
\bibitem [{\citenamefont {Werwi{\'n}ski}\ \emph {et~al.}(2018)\citenamefont
  {Werwi{\'n}ski}, \citenamefont {Edstr{\"o}m}, \citenamefont {Rusz},
  \citenamefont {Hedlund}, \citenamefont {Gunnarsson}, \citenamefont
  {Svedlindh}, \citenamefont {Cedervall},\ and\ \citenamefont
  {Sahlberg}}]{werwinski_magnetocrystalline_2018}%
  \BibitemOpen
  \bibfield  {author} {\bibinfo {author} {\bibfnamefont {M.}~\bibnamefont
  {Werwi{\'n}ski}}, \bibinfo {author} {\bibfnamefont {A.}~\bibnamefont
  {Edstr{\"o}m}}, \bibinfo {author} {\bibfnamefont {J.}~\bibnamefont {Rusz}},
  \bibinfo {author} {\bibfnamefont {D.}~\bibnamefont {Hedlund}}, \bibinfo
  {author} {\bibfnamefont {K.}~\bibnamefont {Gunnarsson}}, \bibinfo {author}
  {\bibfnamefont {P.}~\bibnamefont {Svedlindh}}, \bibinfo {author}
  {\bibfnamefont {J.}~\bibnamefont {Cedervall}},\ and\ \bibinfo {author}
  {\bibfnamefont {M.}~\bibnamefont {Sahlberg}},\ }\bibfield  {title} {\bibinfo
  {title} {Magnetocrystalline anisotropy of
  {{Fe}}{\textsubscript{5}}{{PB}}{\textsubscript{2}} and its alloys with {{Co}}
  and {\emph{5d}} elements: {{A}} combined first-principles and experimental
  study},\ }\href {https://doi.org/10.1103/PhysRevB.98.214431} {\bibfield
  {journal} {\bibinfo  {journal} {Phys. Rev. B}\ }\textbf {\bibinfo {volume}
  {98}},\ \bibinfo {pages} {214431} (\bibinfo {year} {2018})}\BibitemShut
  {NoStop}%
\bibitem [{\citenamefont {K{\"o}rmann}\ \emph {et~al.}(2009)\citenamefont
  {K{\"o}rmann}, \citenamefont {Dick}, \citenamefont {Hickel},\ and\
  \citenamefont {Neugebauer}}]{kormann_pressure_2009}%
  \BibitemOpen
  \bibfield  {author} {\bibinfo {author} {\bibfnamefont {F.}~\bibnamefont
  {K{\"o}rmann}}, \bibinfo {author} {\bibfnamefont {A.}~\bibnamefont {Dick}},
  \bibinfo {author} {\bibfnamefont {T.}~\bibnamefont {Hickel}},\ and\ \bibinfo
  {author} {\bibfnamefont {J.}~\bibnamefont {Neugebauer}},\ }\bibfield  {title}
  {\bibinfo {title} {Pressure dependence of the {{Curie}} temperature in bcc
  iron studied by ab initio simulations},\ }\href
  {https://doi.org/10.1103/PhysRevB.79.184406} {\bibfield  {journal} {\bibinfo
  {journal} {Phys. Rev. B}\ }\textbf {\bibinfo {volume} {79}},\ \bibinfo
  {pages} {184406} (\bibinfo {year} {2009})}\BibitemShut {NoStop}%
\bibitem [{\citenamefont {Matyunina}\ \emph {et~al.}(2018)\citenamefont
  {Matyunina}, \citenamefont {Zagrebin}, \citenamefont {Sokolovskiy},\ and\
  \citenamefont {Buchelnikov}}]{matyunina_ab_2018}%
  \BibitemOpen
  \bibfield  {author} {\bibinfo {author} {\bibfnamefont {M.}~\bibnamefont
  {Matyunina}}, \bibinfo {author} {\bibfnamefont {M.}~\bibnamefont {Zagrebin}},
  \bibinfo {author} {\bibfnamefont {V.}~\bibnamefont {Sokolovskiy}},\ and\
  \bibinfo {author} {\bibfnamefont {V.}~\bibnamefont {Buchelnikov}},\
  }\bibfield  {title} {\bibinfo {title} {Ab initio study of magnetic and
  structural properties of {{Fe-Ga}} alloys},\ }\href
  {https://doi.org/10.1051/epjconf/201818504013} {\bibfield  {journal}
  {\bibinfo  {journal} {EPJ Web Conf.}\ }\textbf {\bibinfo {volume} {185}},\
  \bibinfo {pages} {04013} (\bibinfo {year} {2018})}\BibitemShut {NoStop}%
\bibitem [{\citenamefont {Romero}\ and\ \citenamefont
  {Verstraete}(2018)}]{romero_one_2018}%
  \BibitemOpen
  \bibfield  {author} {\bibinfo {author} {\bibfnamefont {A.~H.}\ \bibnamefont
  {Romero}}\ and\ \bibinfo {author} {\bibfnamefont {M.~J.}\ \bibnamefont
  {Verstraete}},\ }\bibfield  {title} {\bibinfo {title} {From one to three,
  exploring the rungs of {{Jacob}}'s ladder in magnetic alloys},\ }\href
  {https://doi.org/10.1140/epjb/e2018-90275-5} {\bibfield  {journal} {\bibinfo
  {journal} {Eur. Phys. J. B}\ }\textbf {\bibinfo {volume} {91}},\ \bibinfo
  {pages} {193} (\bibinfo {year} {2018})}\BibitemShut {NoStop}%
\bibitem [{\citenamefont {Lee}\ \emph {et~al.}(2021)\citenamefont {Lee},
  \citenamefont {Ryu}, \citenamefont {Kim}, \citenamefont {Byeon},
  \citenamefont {Jeong}, \citenamefont {Lee}, \citenamefont {Hong},
  \citenamefont {Cho}, \citenamefont {Lee}, \citenamefont {Park},\ and\
  \citenamefont {Jeen}}]{lee_large_2021}%
  \BibitemOpen
  \bibfield  {author} {\bibinfo {author} {\bibfnamefont {J.}~\bibnamefont
  {Lee}}, \bibinfo {author} {\bibfnamefont {S.}~\bibnamefont {Ryu}}, \bibinfo
  {author} {\bibfnamefont {I.}~\bibnamefont {Kim}}, \bibinfo {author}
  {\bibfnamefont {M.}~\bibnamefont {Byeon}}, \bibinfo {author} {\bibfnamefont
  {M.-H.}\ \bibnamefont {Jeong}}, \bibinfo {author} {\bibfnamefont {J.~S.}\
  \bibnamefont {Lee}}, \bibinfo {author} {\bibfnamefont {T.~E.}\ \bibnamefont
  {Hong}}, \bibinfo {author} {\bibfnamefont {J.}~\bibnamefont {Cho}}, \bibinfo
  {author} {\bibfnamefont {J.}~\bibnamefont {Lee}}, \bibinfo {author}
  {\bibfnamefont {J.~K.}\ \bibnamefont {Park}},\ and\ \bibinfo {author}
  {\bibfnamefont {H.}~\bibnamefont {Jeen}},\ }\bibfield  {title} {\bibinfo
  {title} {Large enhancement of magnetic moment in nitridated
  {{CeFe}}{\textsubscript{12}}},\ }\href
  {https://doi.org/10.1016/j.jallcom.2021.161245} {\bibfield  {journal}
  {\bibinfo  {journal} {J. Alloys Compd.}\ }\textbf {\bibinfo {volume} {886}},\
  \bibinfo {pages} {161245} (\bibinfo {year} {2021})}\BibitemShut {NoStop}%
\bibitem [{\citenamefont {Schwarz}\ and\ \citenamefont
  {Mohn}(1984)}]{schwarz_itinerant_1984}%
  \BibitemOpen
  \bibfield  {author} {\bibinfo {author} {\bibfnamefont {K.}~\bibnamefont
  {Schwarz}}\ and\ \bibinfo {author} {\bibfnamefont {P.}~\bibnamefont {Mohn}},\
  }\bibfield  {title} {\bibinfo {title} {Itinerant metamagnetism in
  {{YCo}}{$_{2}$}},\ }\href {https://doi.org/10.1088/0305-4608/14/7/008}
  {\bibfield  {journal} {\bibinfo  {journal} {J. Phys. F: Met. Phys.}\ }\textbf
  {\bibinfo {volume} {14}},\ \bibinfo {pages} {L129} (\bibinfo {year}
  {1984})}\BibitemShut {NoStop}%
\bibitem [{\citenamefont {Momma}\ and\ \citenamefont
  {Izumi}(2008)}]{momma_vesta_2008}%
  \BibitemOpen
  \bibfield  {author} {\bibinfo {author} {\bibfnamefont {K.}~\bibnamefont
  {Momma}}\ and\ \bibinfo {author} {\bibfnamefont {F.}~\bibnamefont {Izumi}},\
  }\bibfield  {title} {\bibinfo {title} {{{{\emph{VESTA}}}} : {{A}}
  three-dimensional visualization system for electronic and structural
  analysis},\ }\href {https://doi.org/10.1107/S0021889808012016} {\bibfield
  {journal} {\bibinfo  {journal} {J. Appl. Crystallogr.}\ }\textbf {\bibinfo
  {volume} {41}},\ \bibinfo {pages} {653} (\bibinfo {year} {2008})}\BibitemShut
  {NoStop}%
\bibitem [{\citenamefont {Nieves}\ \emph {et~al.}(2019)\citenamefont {Nieves},
  \citenamefont {Arapan}, \citenamefont {{Maudes-Raedo}}, \citenamefont
  {{Marticorena-S{\'a}nchez}}, \citenamefont {Del~Br{\'i}o}, \citenamefont
  {Kovacs}, \citenamefont {{Echevarria-Bonet}}, \citenamefont {Salazar},
  \citenamefont {Weischenberg}, \citenamefont {Zhang}, \citenamefont
  {Vekilova}, \citenamefont {{Serrano-L{\'o}pez}}, \citenamefont {Barandiaran},
  \citenamefont {Skokov}, \citenamefont {Gutfleisch}, \citenamefont {Eriksson},
  \citenamefont {Herper}, \citenamefont {Schrefl},\ and\ \citenamefont
  {{Cuesta-L{\'o}pez}}}]{nieves_database_2019}%
  \BibitemOpen
  \bibfield  {author} {\bibinfo {author} {\bibfnamefont {P.}~\bibnamefont
  {Nieves}}, \bibinfo {author} {\bibfnamefont {S.}~\bibnamefont {Arapan}},
  \bibinfo {author} {\bibfnamefont {J.}~\bibnamefont {{Maudes-Raedo}}},
  \bibinfo {author} {\bibfnamefont {R.}~\bibnamefont
  {{Marticorena-S{\'a}nchez}}}, \bibinfo {author} {\bibfnamefont {N.~L.}\
  \bibnamefont {Del~Br{\'i}o}}, \bibinfo {author} {\bibfnamefont
  {A.}~\bibnamefont {Kovacs}}, \bibinfo {author} {\bibfnamefont
  {C.}~\bibnamefont {{Echevarria-Bonet}}}, \bibinfo {author} {\bibfnamefont
  {D.}~\bibnamefont {Salazar}}, \bibinfo {author} {\bibfnamefont
  {J.}~\bibnamefont {Weischenberg}}, \bibinfo {author} {\bibfnamefont
  {H.}~\bibnamefont {Zhang}}, \bibinfo {author} {\bibfnamefont {O.~Y.}\
  \bibnamefont {Vekilova}}, \bibinfo {author} {\bibfnamefont {R.}~\bibnamefont
  {{Serrano-L{\'o}pez}}}, \bibinfo {author} {\bibfnamefont {J.~M.}\
  \bibnamefont {Barandiaran}}, \bibinfo {author} {\bibfnamefont
  {K.}~\bibnamefont {Skokov}}, \bibinfo {author} {\bibfnamefont
  {O.}~\bibnamefont {Gutfleisch}}, \bibinfo {author} {\bibfnamefont
  {O.}~\bibnamefont {Eriksson}}, \bibinfo {author} {\bibfnamefont {H.~C.}\
  \bibnamefont {Herper}}, \bibinfo {author} {\bibfnamefont {T.}~\bibnamefont
  {Schrefl}},\ and\ \bibinfo {author} {\bibfnamefont {S.}~\bibnamefont
  {{Cuesta-L{\'o}pez}}},\ }\bibfield  {title} {\bibinfo {title} {Database of
  novel magnetic materials for high-performance permanent magnet development},\
  }\href {https://doi.org/10.1016/j.commatsci.2019.06.007} {\bibfield
  {journal} {\bibinfo  {journal} {Comput. Mater. Sci.}\ }\textbf {\bibinfo
  {volume} {168}},\ \bibinfo {pages} {188} (\bibinfo {year}
  {2019})}\BibitemShut {NoStop}%
\bibitem [{\citenamefont {Burkert}\ \emph {et~al.}(2004)\citenamefont
  {Burkert}, \citenamefont {Nordstr{\"o}m}, \citenamefont {Eriksson},\ and\
  \citenamefont {Heinonen}}]{burkert_giant_2004-1}%
  \BibitemOpen
  \bibfield  {author} {\bibinfo {author} {\bibfnamefont {T.}~\bibnamefont
  {Burkert}}, \bibinfo {author} {\bibfnamefont {L.}~\bibnamefont
  {Nordstr{\"o}m}}, \bibinfo {author} {\bibfnamefont {O.}~\bibnamefont
  {Eriksson}},\ and\ \bibinfo {author} {\bibfnamefont {O.}~\bibnamefont
  {Heinonen}},\ }\bibfield  {title} {\bibinfo {title} {Giant {{Magnetic
  Anisotropy}} in {{Tetragonal FeCo Alloys}}},\ }\href
  {https://doi.org/10.1103/PhysRevLett.93.027203} {\bibfield  {journal}
  {\bibinfo  {journal} {Phys. Rev. Lett.}\ }\textbf {\bibinfo {volume} {93}},\
  \bibinfo {pages} {027203} (\bibinfo {year} {2004})}\BibitemShut {NoStop}%
\bibitem [{\citenamefont {Galler}\ \emph {et~al.}(2021)\citenamefont {Galler},
  \citenamefont {Ener}, \citenamefont {Maccari}, \citenamefont {Dirba},
  \citenamefont {Skokov}, \citenamefont {Gutfleisch}, \citenamefont
  {Biermann},\ and\ \citenamefont {Pourovskii}}]{galler_intrinsically_2020}%
  \BibitemOpen
  \bibfield  {author} {\bibinfo {author} {\bibfnamefont {A.}~\bibnamefont
  {Galler}}, \bibinfo {author} {\bibfnamefont {S.}~\bibnamefont {Ener}},
  \bibinfo {author} {\bibfnamefont {F.}~\bibnamefont {Maccari}}, \bibinfo
  {author} {\bibfnamefont {I.}~\bibnamefont {Dirba}}, \bibinfo {author}
  {\bibfnamefont {K.~P.}\ \bibnamefont {Skokov}}, \bibinfo {author}
  {\bibfnamefont {O.}~\bibnamefont {Gutfleisch}}, \bibinfo {author}
  {\bibfnamefont {S.}~\bibnamefont {Biermann}},\ and\ \bibinfo {author}
  {\bibfnamefont {L.~V.}\ \bibnamefont {Pourovskii}},\ }\bibfield  {title}
  {\bibinfo {title} {Intrinsically weak magnetic anisotropy of cerium in
  potential hard-magnetic intermetallics},\ }\href
  {https://doi.org/10.1038/s41535-020-00301-6} {\bibfield  {journal} {\bibinfo
  {journal} {npj Quantum Mater.}\ }\textbf {\bibinfo {volume} {6}},\ \bibinfo
  {pages} {2} (\bibinfo {year} {2021})}\BibitemShut {NoStop}%
\bibitem [{\citenamefont {Mart{\'i}nez~S{\'a}nchez}\ \emph
  {et~al.}(2020)\citenamefont {Mart{\'i}nez~S{\'a}nchez}, \citenamefont
  {Alfonso}, \citenamefont {Hernandez}, \citenamefont {Jaramillo},\ and\
  \citenamefont {Alc{\'a}zar}}]{martinez_sanchez_effect_2020}%
  \BibitemOpen
  \bibfield  {author} {\bibinfo {author} {\bibfnamefont {H.}~\bibnamefont
  {Mart{\'i}nez~S{\'a}nchez}}, \bibinfo {author} {\bibfnamefont {L.~E.~Z.}\
  \bibnamefont {Alfonso}}, \bibinfo {author} {\bibfnamefont {J.~S.~T.}\
  \bibnamefont {Hernandez}}, \bibinfo {author} {\bibfnamefont {D.~S.}\
  \bibnamefont {Jaramillo}},\ and\ \bibinfo {author} {\bibfnamefont {G.~A.~P.}\
  \bibnamefont {Alc{\'a}zar}},\ }\bibfield  {title} {\bibinfo {title} {Effect
  of {{Nitrogenation}} on the {{Intrinsic Magnetic Properties}} of the
  {{Compounds}}
  ({{Nd}}{\textsubscript{1-x}}{{Ce}}{\textsubscript{x}}){\textsubscript{1.1}}{{Fe}}{\textsubscript{10}}{{CoTi}}},\
  }\href {https://doi.org/10.1109/TMAG.2020.3020094} {\bibfield  {journal}
  {\bibinfo  {journal} {IEEE Trans. Magn.}\ }\textbf {\bibinfo {volume} {56}},\
  \bibinfo {pages} {1} (\bibinfo {year} {2020})}\BibitemShut {NoStop}%
\bibitem [{\citenamefont {Dasmahapatra}\ \emph {et~al.}(2021)\citenamefont
  {Dasmahapatra}, \citenamefont {{Martinez-Casado}}, \citenamefont
  {{Romero-Mu{\~n}iz}}, \citenamefont {Sgroi}, \citenamefont {Ferrari},\ and\
  \citenamefont {Maschio}}]{dasmahapatra_doping_2021}%
  \BibitemOpen
  \bibfield  {author} {\bibinfo {author} {\bibfnamefont {A.}~\bibnamefont
  {Dasmahapatra}}, \bibinfo {author} {\bibfnamefont {R.}~\bibnamefont
  {{Martinez-Casado}}}, \bibinfo {author} {\bibfnamefont {C.}~\bibnamefont
  {{Romero-Mu{\~n}iz}}}, \bibinfo {author} {\bibfnamefont {M.~F.}\ \bibnamefont
  {Sgroi}}, \bibinfo {author} {\bibfnamefont {A.~M.}\ \bibnamefont {Ferrari}},\
  and\ \bibinfo {author} {\bibfnamefont {L.}~\bibnamefont {Maschio}},\
  }\bibfield  {title} {\bibinfo {title} {Doping the permanent magnet
  {{CeFe}}{\textsubscript{11}}{{Ti}} with {{Co}} and {{Ni}} using ab-initio
  density functional methods},\ }\href
  {https://doi.org/10.1016/j.physb.2021.413241} {\bibfield  {journal} {\bibinfo
   {journal} {Physica B: Condensed Matter}\ }\textbf {\bibinfo {volume}
  {620}},\ \bibinfo {pages} {413241} (\bibinfo {year} {2021})}\BibitemShut
  {NoStop}%
\bibitem [{\citenamefont {Harashima}\ \emph {et~al.}(2016)\citenamefont
  {Harashima}, \citenamefont {Terakura}, \citenamefont {Kino}, \citenamefont
  {Ishibashi},\ and\ \citenamefont {Miyake}}]{harashima_first-principles_2016}%
  \BibitemOpen
  \bibfield  {author} {\bibinfo {author} {\bibfnamefont {Y.}~\bibnamefont
  {Harashima}}, \bibinfo {author} {\bibfnamefont {K.}~\bibnamefont {Terakura}},
  \bibinfo {author} {\bibfnamefont {H.}~\bibnamefont {Kino}}, \bibinfo {author}
  {\bibfnamefont {S.}~\bibnamefont {Ishibashi}},\ and\ \bibinfo {author}
  {\bibfnamefont {T.}~\bibnamefont {Miyake}},\ }\bibfield  {title} {\bibinfo
  {title} {First-principles study on stability and magnetism of
  {{NdFe}}{\textsubscript{11}}{{M}} and {{NdFe}}{\textsubscript{11}}{{MN}} for
  {{M}}\,=\,{{Ti}}, {{V}}, {{Cr}}, {{Mn}}, {{Fe}}, {{Co}}, {{Ni}}, {{Cu}},
  {{Zn}}},\ }\href {https://doi.org/10.1063/1.4968798} {\bibfield  {journal}
  {\bibinfo  {journal} {J. Appl. Phys.}\ }\textbf {\bibinfo {volume} {120}},\
  \bibinfo {pages} {203904} (\bibinfo {year} {2016})}\BibitemShut {NoStop}%
\bibitem [{\citenamefont {Akai}(1988)}]{akai_nuclear_nodate}%
  \BibitemOpen
  \bibfield  {author} {\bibinfo {author} {\bibfnamefont {H.}~\bibnamefont
  {Akai}},\ }\bibfield  {title} {\bibinfo {title} {Nuclear spin-lattice
  relaxation of impurities in ferromagnetic iron},\ }\href
  {https://doi.org/10.1007/BF02398306} {\bibfield  {journal} {\bibinfo
  {journal} {Hyperfine Interact.}\ }\textbf {\bibinfo {volume} {43}},\ \bibinfo
  {pages} {253} (\bibinfo {year} {1988})}\BibitemShut {NoStop}%
\bibitem [{\citenamefont {Dederichs}\ \emph {et~al.}(1991)\citenamefont
  {Dederichs}, \citenamefont {Zeller}, \citenamefont {Akai},\ and\
  \citenamefont {Ebert}}]{dederichs_ab-initio_1991-1}%
  \BibitemOpen
  \bibfield  {author} {\bibinfo {author} {\bibfnamefont {P.~H.}\ \bibnamefont
  {Dederichs}}, \bibinfo {author} {\bibfnamefont {R.}~\bibnamefont {Zeller}},
  \bibinfo {author} {\bibfnamefont {H.}~\bibnamefont {Akai}},\ and\ \bibinfo
  {author} {\bibfnamefont {H.}~\bibnamefont {Ebert}},\ }\bibfield  {title}
  {\bibinfo {title} {Ab-initio calculations of the electronic structure of
  impurities and alloys of ferromagnetic transition metals},\ }\href
  {https://doi.org/10.1016/0304-8853(91)90823-S} {\bibfield  {journal}
  {\bibinfo  {journal} {J. Magn. Magn. Mater.}\ }\textbf {\bibinfo {volume}
  {100}},\ \bibinfo {pages} {241} (\bibinfo {year} {1991})}\BibitemShut
  {NoStop}%
\bibitem [{\citenamefont {Wienke}\ \emph {et~al.}(1991)\citenamefont {Wienke},
  \citenamefont {Sch{\"u}tz},\ and\ \citenamefont
  {Ebert}}]{wienke_determination_1991-1}%
  \BibitemOpen
  \bibfield  {author} {\bibinfo {author} {\bibfnamefont {R.}~\bibnamefont
  {Wienke}}, \bibinfo {author} {\bibfnamefont {G.}~\bibnamefont {Sch{\"u}tz}},\
  and\ \bibinfo {author} {\bibfnamefont {H.}~\bibnamefont {Ebert}},\ }\bibfield
   {title} {\bibinfo {title} {Determination of local magnetic moments of 5d
  impurities in {{Fe}} detected via spin-dependent absorption},\ }\href
  {https://doi.org/10.1063/1.348786} {\bibfield  {journal} {\bibinfo  {journal}
  {J. Appl. Phys.}\ }\textbf {\bibinfo {volume} {69}},\ \bibinfo {pages} {6147}
  (\bibinfo {year} {1991})}\BibitemShut {NoStop}%
\bibitem [{\citenamefont {Bartashevich}\ \emph {et~al.}(1994)\citenamefont
  {Bartashevich}, \citenamefont {Goto}, \citenamefont {Radwanski},\ and\
  \citenamefont {Korolyov}}]{bartashevich_magnetic_1994}%
  \BibitemOpen
  \bibfield  {author} {\bibinfo {author} {\bibfnamefont {M.~I.}\ \bibnamefont
  {Bartashevich}}, \bibinfo {author} {\bibfnamefont {T.}~\bibnamefont {Goto}},
  \bibinfo {author} {\bibfnamefont {R.~J.}\ \bibnamefont {Radwanski}},\ and\
  \bibinfo {author} {\bibfnamefont {A.~V.}\ \bibnamefont {Korolyov}},\
  }\bibfield  {title} {\bibinfo {title} {Magnetic anisotropy and high-field
  magnetization process of {{CeCo}}{\textsubscript{5}}},\ }\href
  {https://doi.org/10.1016/0304-8853(94)90010-8} {\bibfield  {journal}
  {\bibinfo  {journal} {J. Magn. Magn. Mater.}\ }\textbf {\bibinfo {volume}
  {131}},\ \bibinfo {pages} {61} (\bibinfo {year} {1994})}\BibitemShut
  {NoStop}%
\bibitem [{\citenamefont {Capehart}\ \emph {et~al.}(1993)\citenamefont
  {Capehart}, \citenamefont {Mishra}, \citenamefont {Meisner}, \citenamefont
  {Fuerst},\ and\ \citenamefont {Herbst}}]{capehart_steric_1993}%
  \BibitemOpen
  \bibfield  {author} {\bibinfo {author} {\bibfnamefont {T.~W.}\ \bibnamefont
  {Capehart}}, \bibinfo {author} {\bibfnamefont {R.~K.}\ \bibnamefont
  {Mishra}}, \bibinfo {author} {\bibfnamefont {G.~P.}\ \bibnamefont {Meisner}},
  \bibinfo {author} {\bibfnamefont {C.~D.}\ \bibnamefont {Fuerst}},\ and\
  \bibinfo {author} {\bibfnamefont {J.~F.}\ \bibnamefont {Herbst}},\ }\bibfield
   {title} {\bibinfo {title} {Steric variation of the cerium valence in
  {{Ce}}{\textsubscript{2}}{{Fe}}{\textsubscript{14}}{{B}} and related
  compounds},\ }\href {https://doi.org/10.1063/1.110075} {\bibfield  {journal}
  {\bibinfo  {journal} {Appl. Phys. Lett.}\ }\textbf {\bibinfo {volume} {63}},\
  \bibinfo {pages} {3642} (\bibinfo {year} {1993})}\BibitemShut {NoStop}%
\end{thebibliography}%
\break   

\end{document}